\documentclass[openacc]{rsproca_new}

\titlehead{Research}

\begin{document}

\title{Post-injection aseismic slip as a mechanism for the delayed triggering of seismicity}

\author{
Alexis S\'aez and Brice Lecampion}

\address{Ecole Polytechnique F\'ed\'erale de Lausanne (EPFL), Institute of Civil Engineering, Gaznat chair on Geo-Energy, CH-1015 Lausanne, Switzerland}

\subject{geophysics, mechanics, civil engineering}


\corres{Alexis S\'aez\\
\email{alexis.saez@epfl.ch}}

\begin{abstract}
Injection-induced aseismic slip plays an important role in a broad range of human-made and natural systems, from the exploitation of geo-resources to the understanding of earthquakes. Recent studies have shed light on how aseismic slip propagates in response to continuous fluid injections. Yet much less is known about the response of faults after the injection of fluids has stopped. In this work, we investigate via an hydro-mechanical model the propagation and ultimate arrest of aseismic slip during the so-called post-injection stage. We show that after shut-in, fault slip propagates in pulse-like mode. The conditions that control the propagation as a pulse and notably when and where the ruptures arrest are fully established. In particular, critically-stressed faults can host rupture pulses that propagate for several orders of magnitude the injection duration and reach up to nearly double the size of the ruptures at the moment of shut-in. We consequently argue that the persistent stressing of increasingly larger rock volumes caused by post-injection aseismic slip is a plausible mechanism for the triggering of post-injection seismicity ---a critical issue in the geo-energy industry. We discuss evidence that supports this mechanism based on documented cases of post-injection-induced seismicity.
\end{abstract}

\begin{fmtext}

\section{Introduction}
\label{introduction}
It has been long recognized that subsurface fluid injections can induce fast fault slip \cite{Healy_Rubey_1968} that radiates detectable elastodynamic waves and also slow, sometimes called aseismic slip \cite{Hamilton_Meehan_1971}, which is generally more difficult to detect by classical monitoring networks. Evidence for injection-induced aseismic slip dates back to the 1960's, when a slow surface fault rupture was causally linked to fluid injection operations of an oil field in LA, {{\parfillskip0pt\par}}
\end{fmtext}


\maketitle
\noindent USA \cite{Hamilton_Meehan_1971}. Since then, an increasing number of studies have inferred the occurrence of aseismic slip episodes as a result of anthropogenic subsurface fluid injections \cite{Scotti_Cornet_1994,Bourouis_Bernard_2007,Wei_Avouac_2015,Eyre_Samsonov_2022}, with recent in-situ experiments of fluid injection into natural fault systems demonstrating by direct measurements of fault deformation that slow slip was systematically the dominant style of fault motion \cite{Guglielmi_Cappa_2015,Cappa_Scuderi_2019}.

Injection-induced aseismic slip is thought to play an important role in seismicity induced by industrial fluid injections, a phenomenon that has become critical in ensuring the sustainable development of unconventional hydrocarbon reservoirs \cite{Bao_Eaton_2016} and deep geothermal resources \cite{Deichmann_Giardini_2009,Ellsworth_Giardini_2019}. It is understood that fluid-driven slow ruptures transmit solid stresses quasi-statically to unstable fault patches and trigger instabilities at distances that can be far from the region affected by the pressurization of pore-fluid \cite{Eyre_Eaton_2019,Bhattacharya_Viesca_2019}. Moreover, a similar mechanism might be also operating behind some natural episodes of seismicity such as seismic swarms and aftershock sequences. In fact, both phenomena are commonly interpreted to be driven by either the diffusion of pore pressure \cite{Parotidis_Shapiro_2005,Miller_Collettini_2004} or the propagation of aseismic slip \cite{Lohman_McGuire_2007,Perfettini_Avouac_2007}, with recent studies suggesting that both mechanisms might be indeed coupled and responsible for the observed spatio-temporal patterns of seismicity \cite{Yukutake_Yoshida_2022,Sirorattanakul_Ross_2022,Ross_Rollins_2017}. Likewise, tectonic tremors and low-frequency earthquakes are often considered to be driven by slow slip events in subduction zones \cite{Rogers_Dragert_2003,Shelly_Beroza_2006}, at depths where systematic evidence of over-pressurized fluids is found \cite{Shelly_Beroza_2006,Kato_Iidaka_2010}. Metamorphic dehydration reactions \cite{Peacock_2009} and fault-valving behavior \cite{Sibson_2020} are the two common candidates to explain not only this inferred over-pressurization, but also the very nature of pore pressure and aseismic slip transients in these zones \cite{Zhu_Allison_2020}.

The seemingly relevant role of fluid-driven aseismic slip in the previous phenomena have motivated the development of physical models that, in recent years, have contributed to a better understanding of the mechanics of this hydro-mechanical problem. Some recent advances include a better notion of how the initial state of stress, the fluid injection parameters, the fault hydraulic properties, and the possible rate-strengthening dependence of rock friction, may affect the dynamics of fluid-driven aseismic slip transients in 2D \cite{Dublanchet_2019,Garagash_2021,Viesca_2021,Yang_Dunham_2021} and 3D media \cite{Saez_Lecampion_2022}. The three-dimensional case is the relevant one for field applications and has been shown to be not only quantitatively but also qualitatively very different from two-dimensional configurations \cite{Saez_Lecampion_2022}. 
One common aspect of all prior studies is that they focus on fluid sources that are continuous (uninterrupted) in time. Yet much less is known about aseismic ruptures after the injection of fluids has stopped. In particular, the conditions that control the further propagation and ultimate arrest of fluid-driven aseismic slip are fairly unknown, despite a recent investigation for a two-dimensional plane-strain configuration \cite{Jacquey_Viesca_2022}. This is of broad interest since, ultimately, any kind of fluid source and accompanying rupture will have to stop. Understanding the mechanics of post-injection aseismic slip in a realistic three-dimensional scenario is thus of major importance. It corresponds to the first goal of this study.

Our second goal is to understand how and to which extent post-injection aseismic slip can be considered as a possible mechanism for the triggering of seismicity after the end of subsurface fluid injections. It is well-known since the Denver earthquakes in the 1960s \cite{Healy_Rubey_1968} that upon shutting off the wells, seismicity might continue to occur. Multiple observations suggest that indeed, it is not rare that the largest events of injection-induced seismic sequences happen during the post-injection stage. Examples of those cases are the 2006 $M_{L}>3$ Basel earthquakes in Switzerland, and the 2017 $M_{w}$ 5.5 Pohang earthquake in South Korea, both causally linked to hydraulic stimulation operations of deep geothermal reservoirs \cite{Deichmann_Giardini_2009,Ellsworth_Giardini_2019}. Also, the 2015 $M_{w}$ 3.9 earthquake in western Canada, that occurred after the end of hydraulic fracturing operations in a hydrocarbon reservoir \cite{Bao_Eaton_2016}. Because the shut-in of the wells does not guarantee the cessation of seismic events, current efforts to manage the seismic risk in the geo-energy industry such as the so-called traffic light systems \cite{Baisch_Koch_2019}, might be subjected to important limitations in their effectiveness. Understanding the physical mechanisms underpinning post-injection seismicity is thus of paramount importance for the successful development of both hydrocarbon and geothermal reservoirs. We therefore aim in this study at understanding the combined effect of pore pressure diffusion and transmission of solid stresses due to aseismic slip in the triggering of post-injection seismicity.

This article is organized as follows. In section \ref{model}, we present the governing equations and scaling analysis of our model, as well as the chosen numerical methods. In section \ref{pore-pressure}, we examine the spatio-temporal evolution of pore pressure after shut-in. In section \ref{circular}, we present the results of post-injection aseismic slip for the particular yet insightful case of circular ruptures. In section \ref{non-circular}, we explore the more general case of non-circular ruptures. In section \ref{discussion}, we elaborate on the triggering mechanism of seismicity due to post-injection aseismic slip and discuss possible evidence for it based on field cases. Finally, in section \ref{conclusion}, we provide some concluding remarks of our work.

\section{Governing equations, scaling analysis and numerical methods}
\label{model}

\subsection{Governing equations}
\label{governing-equations}
\begin{figure}
\centering
  \includegraphics[width=10cm]{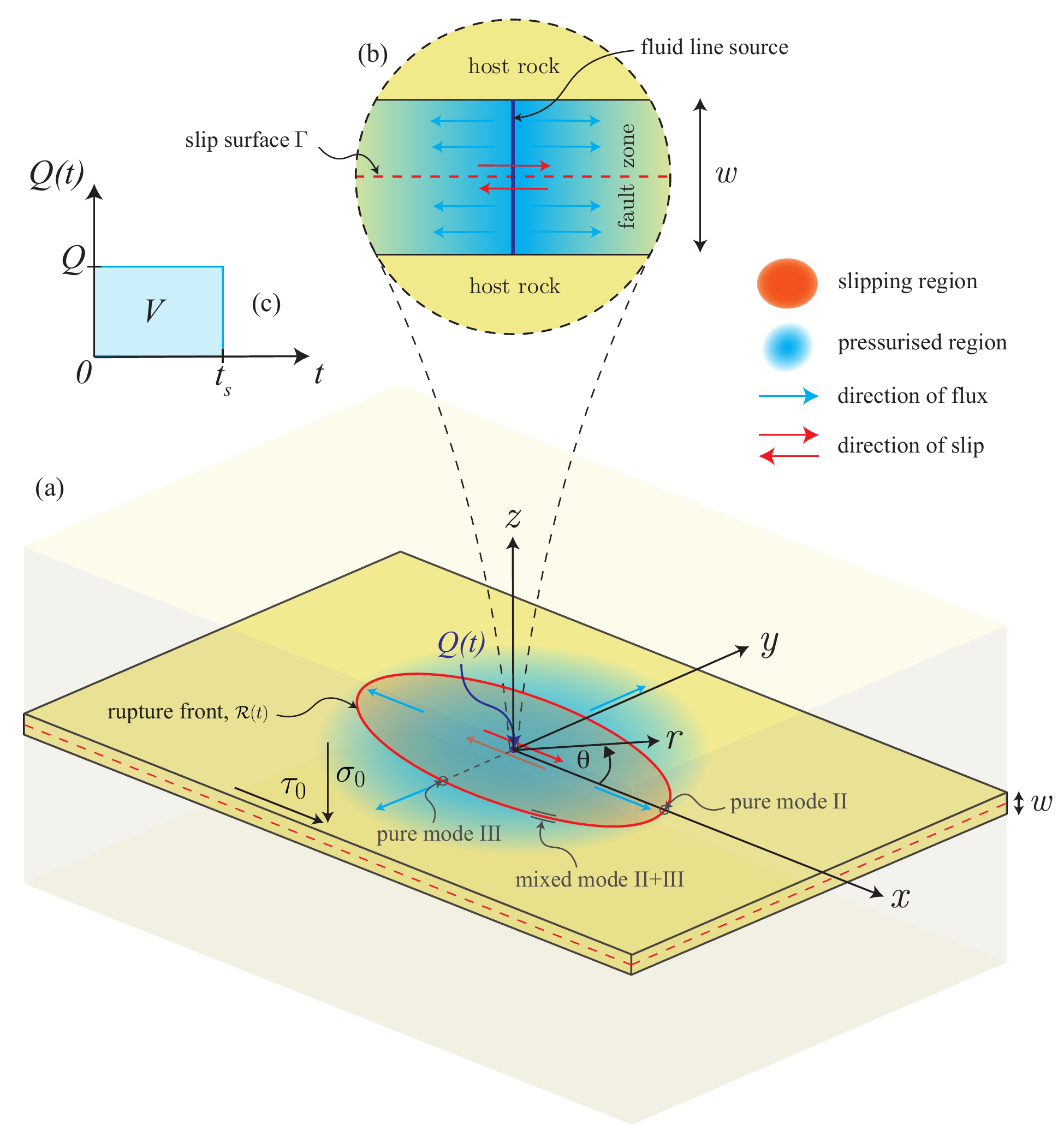}
  \caption{Model schematics. (a) and (b), fluid is injected into a permeable fault zone of width $w$ via a line-source that crosses the entire fault zone width. The fault is planar and embedded in an unbounded linearly elastic impermeable host rock. The initial stress tensor is uniform. (c) Fluid is injected at a constant rate $Q$ until $t=t_s$ (the shut-in time) at which the injection is instantaneously stopped.}
  \label{fig:model-schematics}
\end{figure}

We consider a planar fault in an infinite, linearly elastic, isotropic, impermeable, and three-dimensional solid, as depicted in figure  \ref{fig:model-schematics}. The initial stress tensor is assumed to be uniform and is characterized by a shear stress $\tau_{0}$ resolved on the fault plane that acts along the \textit{x} direction and a total normal stress $\sigma_{0}$ (acting along the \textit{z} direction). We assume a uniform initial pore pressure field of magnitude $p_{0}$ that is perturbed by the sudden injection of fluids into a poroelastic fault zone of width $w$. We model such fluid injection via a line-source that is located along the \textit{z} axis and crosses the entire fault width. The fault zone permeability $k$ and storage coefficient $S$ are assumed to be uniform and constant. As a result, fluid flow is axisymmetric with regard to the \textit{z} axis and occurs only within the fault zone. We further assume that the shear modulus $\mu$ and Poisson's ratio $\nu$ of the poroelastic fault zone and impermeable elastic host rock are the same. Under such conditions, the displacement field induced by the fluid injection is irrotational and the pore pressure diffusion equation of poroelasticity reduces to its uncoupled version \cite{Marck_Savitski_2015}, $\partial p/\partial t=\alpha\nabla^{2}p$, where $\alpha=k/S\eta$ is the fault hydraulic diffusivity, with $\eta$ the fluid dynamic viscosity.

Owing to the planarity of the fault plane $\Gamma$ and the unidirectionality of the initial shear stress resolved on the fault plane, fluid flow induces fault slip $\delta$ and changes in the shear stress $\tau$ that are both characterized by a uniform direction along the \textit{x} axis. By neglecting any poroelastic coupling within the fault zone upon activation of slip, the quasi-static elastic equilibrium that relates fault slip to the shear stress can be written as the following boundary integral equation along the \textit{x} axis,
\begin{equation}\label{eq:momentum}
    \tau(x,y,t)=\tau_{0}+\int_{\Gamma}K(x-\xi,y-\zeta;\mu,\nu)\delta(\xi,\zeta,t)\textrm{d}\xi\textrm{d}\zeta,
\end{equation}
where $K$ is the hypersingular (of order $1/r^{3}$) elastostatic traction fundamental solution (see \cite{Hills_Kelly_1996} for example).

The fault slip surface $\Gamma$ is assumed to obey a Mohr-Coulomb shear failure criterion with a constant friction coefficient $f$ and no cohesion expressed as the following inequality,
\begin{equation}\label{eq:MC-criterion}
    \left|\tau(x,y,t)\right|\leq f\left(\sigma_{0}^{\prime}-\Delta p(r,t)\right),
\end{equation}
where $\sigma_{0}^{\prime}=\sigma_{0}-p_{0}$ is the initial effective normal stress, and $\Delta p(r,t)=p(r,t)-p_{0}$ is the axisymmetric pore pressure perturbation, with $r$ the radial coordinate with its origin at the injection location. 

The injection of fluid starts at $t=0$ and consists of two subsequent stages: a continuous injection stage characterized by a constant rate of injection $Q$, and a post-injection stage in which the injection rate drops instantaneously to zero at $t=t_{s}$ (see figure \ref{fig:model-schematics}c). Hereafter, we refer to $t_{s}$ as the shut-in time. The solution of the linear diffusion equation before shut-in (when $0<t\leq t_{s}$) is known as $\Delta p(r,t)=\Delta p_{*}E_{1}\left(r^{2}/4\alpha t\right)$ (section 10.4, eq. 5, \cite{Carslaw_Jaeger_1959}), where
\begin{equation}\label{eq:p*}
    \Delta p_{*}=\frac{Q\eta}{4\pi kw }
\end{equation}
is the intensity of the injection with units of pressure, and $E_{1}\left(x\right)=\int_{x}^{\infty}\left(e^{-x\xi}/\xi\right)\textrm{d}\xi$ is the exponential integral function. The solution after shut-in (when $t>t_{s}$), that is the stage of interest in the present study, is obtained simply by superposition as
\begin{equation}\label{eq:p-solution}
    \Delta p(r,t)=\Delta p_{*}\left\{ E_{1}\left(\frac{r^{2}}{4\alpha t}\right)-E_{1}\left(\frac{r^{2}}{4\alpha\left(t-t_{s}\right)}\right)\right\}.
\end{equation}
Equations \eqref{eq:momentum}, \eqref{eq:MC-criterion} and \eqref{eq:p-solution} constitute a complete system of equations to solve for the spatio-temporal evolution of fault slip $\delta(x,y,t)$ and, particularly, the moving boundary representing the slipping region $\mathcal{S}(t)$. The slipping patch $\mathcal{S}(t)$ may be defined mathematically as the region in which the equality of equation \eqref{eq:MC-criterion} holds: $\mathcal{S}(t)=\left\{ \left(x,y\right)\in\Gamma:\left|\tau(x,y,t)\right|=f\left(\sigma_{0}^{\prime}-\Delta p(r,t)\right)\right\}$. In general, we are also interested in determining the frontier of $\mathcal{S}(t)$ that corresponds to the rupture front $\mathcal{R}(t)$ when crack-like propagation occurs, as depicted in figure \ref{fig:model-schematics}a. The initial conditions are naturally taken as $\delta(x,y,0)=0$ and $\dot{\delta}(x,y,0)=0$ (fault initially at rest and fully locked).

\subsection{Scaling analysis and limiting regimes}
The shut-in of the injection provides the natural characteristic timescale of the problem, the shut-in time $t_{s}$. This latter introduces, via the solution of the pore-pressure diffusion equation \eqref{eq:p-solution}, a characteristic diffusion lengthscale $\sqrt{4\alpha t_{s}}$. Alternatively, one may choose to scale the spatial coordinates by a characteristic rupture lengthscale at the moment of shut-in, say $R_{s}^{*}$. We normalize the pore pressure perturbation $\Delta p(r,t)$, equation \eqref{eq:p-solution}, by the intensity of the injection $\Delta p_{*}$. Introducing the foregoing characteristic scales in the Mohr-Coulomb shear failure criterion, equation \eqref{eq:MC-criterion}, gives the normalized shear stress, $\left(\tau-f\sigma_{0}^{\prime}\right)/f\Delta p_{*}$, which in turn is introduced in the balance of momentum, equation \eqref{eq:momentum}, allowing us to close the scaling of the problem by normalizing the fault slip by $\left(f\Delta p_{*}/\mu\right)\sqrt{4\alpha t_{s}}$, or alternatively by $\left(f\Delta p_{*}/\mu\right)R_{s}^{*}$ if the characteristic rupture lengthscale is chosen for the spatial scale. 

It can be shown by using the previous scales (see electronic
supplementary material) that the model depends in addition to dimensionless space $\overrightarrow{x}/\sqrt{4\alpha t_{s}}$ and time $t/t_s$ on one single dimensionless number
\begin{equation}
\label{eq:T-parameter}
    \mathcal{T}=\frac{f\sigma_{0}^{\prime}-\tau_{0}}{f \Delta p_{*}},
\end{equation}
and the Poisson's ratio $\nu$.

The so-called stress-injection parameter $\mathcal{T}$ is identical to the one presented recently by S\'aez \textit{et al}. \cite{Saez_Lecampion_2022} for the problem of continuous injection (for times $t\leq t_{s}$). $\mathcal{T}$ is defined as the ratio between the amount of stress necessary to activate fault slip $f\sigma_{0}^{\prime}-\tau_{0}$, and $f\Delta p_{*}$ which is an approximate measure of the reduction of fault shear strength near the injection point due to fluid injection. $\Delta p_{*}$ is the intensity of the injection defined previously in equation \eqref{eq:p*}. $\mathcal{T}$ can vary between $0$ and $+\infty$ and its limiting values are associated with two important end-member scenarios. When $\mathcal{T}\ll1$, the condition $f\sigma_{0}^{\prime}-\tau_{0}\ll f\Delta p_{*}$ means that the fault is “critically stressed” with regard to the reduction of fault shear strength due to fluid injection (a small amount of stress is necessary to activate slip) . On the other hand, when $\mathcal{T}\gg 1$, the condition $f\Delta p_{*}\ll f\sigma_{0}^{\prime}-\tau_{0}$ must be satisfied, so that the fault is “marginally pressurized” with regard to amount of stress necessary to activate fault slip. We will make extensive use of the terms critically-stressed fault/regime and marginally-pressurized fault/regime (the latter adopted first in \cite{Garagash_Germanovich_2012}) to refer to these limiting cases throughout this article. 

As shown recently by S\'aez \textit{et al}. \cite{Saez_Lecampion_2022}, an important aspect of the foregoing two limiting regimes is that they follow different scales during the stage of continuous injection, which is particularly valid in the post-injection problem right at the moment of shut-in $t_{s}$. Indeed, the proper slip scale $\delta_*$ in the critically-stressed limit comes from choosing $\sqrt{4\alpha t_{s}}$ as the characteristic lengthscale, leading to
\begin{equation}
    \delta\sim\delta_*=\frac{f\Delta p_{*}}{\mu}\sqrt{4\alpha t_{s}}.
    \label{eq:slip-scale}
\end{equation}
Here, the $\sim$ symbol means "scales as" or "is of the order of". The proper scale for the slip rate follows from differentiation of the slip scale with respect to time,
\begin{equation}
    v\sim v_*=\frac{f\Delta p_{*}}{\mu}\sqrt{\frac{\alpha}{t_{s}}}.
    \label{eq:slip-rate-scale}
\end{equation}
The slip and slip rate scales of the marginally-pressurized limit can be obtained likewise but choosing the characteristic rupture scale $R_{s}^{*}$ as the spatial lengthscale of the problem, i.e., $\delta\sim\left(f\Delta p_{*}/\mu\right)R_{s}^{*}$ and $v\sim\left(f\Delta p_{*}/\mu\right)R_{s}^{*}/t_{s}$. Nonetheless, the latter scales will be seen later to be somewhat irrelevant in the post-injection stage, because marginally-pressurized faults will host ruptures that after shut-in arrest almost immediately. It seems nevertheless worthwhile noting that this type of scaling relation is the one expected when the shear stress distribution within a quasi-static crack (rupture) is relatively uniform \cite{Kanamori_Anderson_1975}. This is approximately the case of marginally-pressurized faults but not the case of critically-stressed faults. In the latter, the shear stress reduction due to fluid injection is highly-localized near the fluid source at the moment of shut-in, approximately equal to a point-force at distances far from $\sqrt{4\alpha t_{s}}$ \cite{Saez_Lecampion_2022}. This is the reason why the proper lengthscale in the critically-stressed limit is the size of the highly-localized equivalent shear load ($\sqrt{4\alpha t_{s}}$) and not the size of the rupture ($R_{s}^{*}$).

Finally, the second dimensionless parameter of the model, the Poisson's ratio $\nu$, is expected to have only an effect on the rupture shape. As shown by S\'aez \textit{et al}. \cite{Saez_Lecampion_2022} for the continuous-injection stage, the Poisson's ratio modifies the aspect ratio of the resulting rupture fronts which become more elongated for increasing values of $\nu$. The position of the elongated rupture fronts is nevertheless determined primarily by the position of circular ruptures, that correspond to a limiting case in which the Poisson's ratio of the solid $\nu$ is zero. The case of circular ruptures is thus particularly insightful and notably simpler, since in that limit, we can take advantage of the axisymmetry property of the mechanical problem \cite{Bhattacharya_Viesca_2019,Saez_Lecampion_2022}.

\subsection{Numerical methods}
\label{numerical-methods}
We solve the governing system of equations via a fully-implicit boundary-element-based solver with an elasto-plastic-like interfacial constitutive law. The details of the numerical method were presented recently by S\'aez \textit{et al}. \cite{Saez_Lecampion_2022}. The only modification in this work, is the use of a triangular boundary element with constant displacement discontinuity, instead of quadratic shape functions as used originally in \cite{Saez_Lecampion_2022}. Furthermore, to solve for the particular case of circular ruptures (when $\nu=0$), we implemented a more efficient axisymmetric version of our solver based on ring-dislocation boundary elements \cite{Gordeliy_Detournay_2011}. The details of our numerical implementations are given in the electronic supplementary material.

\section{Pore-pressure diffusion: the pore-pressure back front}
\label{pore-pressure}
\begin{figure}
\centering
  \includegraphics[width=13cm]{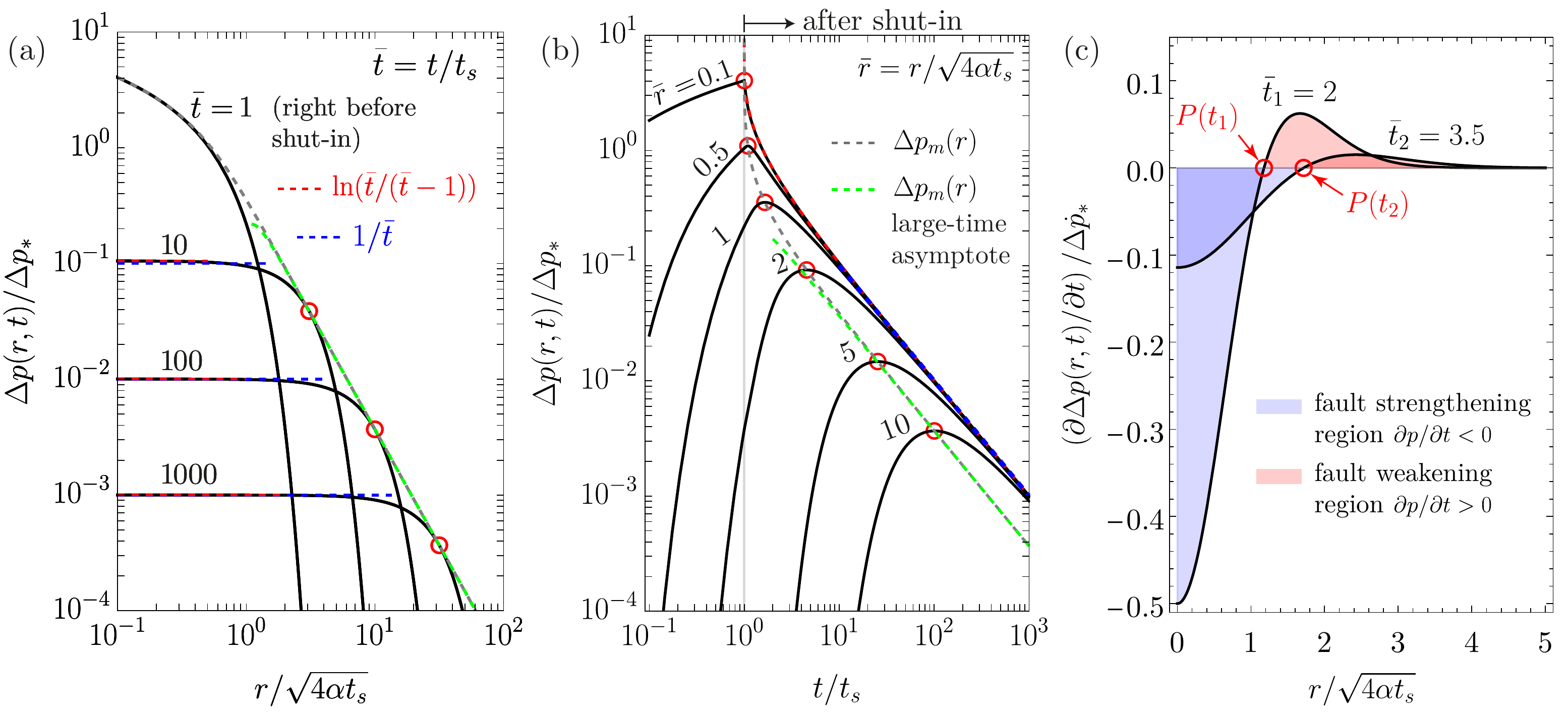}
  \caption{Pore pressure diffusion after shut-in, asymptotic behaviors, and the pore-pressure back front. (a) Normalized spatial distribution of pore pressure at different dimensionless times. Red dashed curves correspond to the asymptotic behavior when $r\ll\sqrt{4\alpha\left(t-t_{s}\right)}$, with the blue dashed curves corresponding to the large-time limit ($t\gg t_s$). (b) Evolution of normalized pore pressure at some fixed dimensionless radial positions. Red circles indicate the instants at which the maximum pressure $\Delta p_{m}$ is reached (when $\partial \Delta p/\partial t=0$). Gray dashed curve is $\Delta p_{m}$ as a function of either $r$ in panel (a) or $t$ in panel (b), with the green dashed curve corresponding to its large-time asymptote, equation \eqref{eq:p-maximum} (c) Spatial distribution of pore pressure rate normalized by $\Delta \dot{p}_*=\Delta p_*/t_s$, at two dimensionless times after shut-in. Red circles indicate the position of the pore-pressure back front $P(t)$, equation \eqref{eq:P-front}, which defines regions where the fault strength is instantaneously increasing ($r<P(t)$) or decreasing ($r>P(t)$).}
  \label{fig:pp-solution}
\end{figure}
The pore-pressure perturbation $\Delta p(r,t)$ is the only external action driving the rupture after shut-in. It is thus essential to examine in detail its spatio-temporal evolution. Various normalized spatial distributions of pore pressure are plotted in figure \ref{fig:pp-solution}a at different dimensionless times. We first observe that pore pressure decreases quickly after the stop of injection near the fluid source in the sense $r\ll\sqrt{4\alpha\left(t-t_{s}\right)}$, asymptotically as $\Delta p(r,t)/\Delta p_{*}\approx\ln\left(\bar{t}/\left(\bar{t}-1\right)\right)$ (red dashed curve), with $\bar{t}=t/t_s$ the dimensionless time. Moreover, at large times ($\bar{t}\gg 1$), it decreases simply as $\Delta p(r,t)/\Delta p_{*}\approx 1/\bar{t}$ (blue dashed curve) for points that are close to the injection now in the sense $r\ll\sqrt{4\alpha t}$. The latter indicates that the region in which the previous asymptotic behavior is approximately satisfied grows diffusively ($\propto \sqrt{\alpha t}$), and has the shape of a plateau in figure \ref{fig:pp-solution}a.

Moreover, figure \ref{fig:pp-solution}a displays that after shut-in the pore pressure keeps increasing away from the injection line. To better understand such process, it may result convenient to look at the temporal evolution of pore pressure at fixed normalized positions over the fault plane. Figure \ref{fig:pp-solution}b displays such a plot. Here, we observe that pore pressure remains increasing for some time after injection stops ($\bar{t}>1$) and eventually reaches a maximum (red circles). After this instant of maximum pressure, the pore pressure decreases following ultimately the same asymptotes of figure \ref{fig:pp-solution}a, meaning that the aforementioned diffusely expanding plateau has reached the corresponding fixed radial positions in figure \ref{fig:pp-solution}b.

The relation between a given position over the fault plane and the time at which the maximum pressure is reached, defines a "back front" of pore pressure, $\mathcal{P}(t)=\left\{ r(t)\in\Gamma:\partial\Delta p(r,t)/\partial t=0\right\}$. Owing to the axisymmetry property of pore pressure diffusion in our model, $\mathcal{P}(t)$ is circular and is thus fully defined by its radius $P(t)$. Differentiating equation \eqref{eq:p-solution} with respect to time and solving for $\partial\Delta p/\partial t=0$ leads to the following expression for the normalized radius of the pore-pressure back front,
\begin{equation}\label{eq:P-front}
    \frac{P(t)}{\sqrt{4\alpha t_{s}}}=\left[\bar{t}\left(\bar{t}-1\right)\ln\left(\frac{\bar{t}}{\bar{t}-1}\right)\right]^{1/2},
\end{equation}
which further reduces to $P(t)/\sqrt{4\alpha t_{s}}\approx\sqrt{\bar{t}}$, or simply $P(t)\approx\sqrt{4\alpha t}$, at large times. Note that the large-time asymptote of $P(t)$ is the same expression than that of the radius of the pore-pressure perturbation front $L(t)=\sqrt{4\alpha t}$ for the continuous-injection stage, albeit each expression describes different processes.

On the other hand, the maximum pore pressure increase undergone historically at a given position $r$, say $\Delta p_{m}(r)$, is obtained by simply evaluating the pore pressure perturbation $\Delta p(r,t)$ at the time $t_{m}$ in which the maximum occurs, i.e., $\Delta p(r,t_{m}(r))$. $t_{m}(r)$ comes from solving for time in \eqref{eq:P-front} with $P(t)=r$. $\Delta p_{m}(r)$ is displayed in figure \ref{fig:pp-solution}a (gray dashed curve) representing the pore pressure back front $\mathcal{P}(t)$ in this plot, which corresponds to the envelope of all curves (for all times) associated with instantaneous spatial distributions of pore pressure. Note that $\Delta p_{m}(r)$ cannot be obtained in closed form since equation \eqref{eq:P-front} is not invertible for time, nevertheless, in the large-time limit, $t_{m}(r)/t_{s}\approx\bar{r}^{2}$, such that 
\begin{equation}\label{eq:p-maximum}
    \Delta p_{m}(r)/\Delta p_{*}\approx E_{1}\left(1\right)-E_{1}\left(\bar{r}^{2}/\left(\bar{r}^{2}-1\right)\right)
\end{equation}
The previous asymptotic approximation is also plotted in figure \ref{fig:pp-solution}a (green dashed curve). Similarly, an expression for $\Delta p_{m}$ as a function of time $t$ may be derived, which is indeed obtainable in closed form as $\Delta p_{m}(t)=\Delta p(P(t),t)$ (gray dashed curve in figure \ref{fig:pp-solution}b). At large times, such expression further simplifies and is equivalent to equation \eqref{eq:p-maximum} but replacing $\bar{r}^{2}$ by $\bar{t}$ (green dashed curve in figure \ref{fig:pp-solution}b).

\section{Pulse-like circular ruptures}
\label{circular}
Circular ruptures occur in the limit of a Poisson's ratio $\nu=0$. As discussed in previous sections, a Poisson's ratio different than zero is expected to affect mainly the shape of the resulting ruptures, and less significantly other relevant quantities of the fault response \cite{Saez_Lecampion_2022}. Circular ruptures are
thus a particularly insightful case, in which the axisymmetry property of the mechanical problem \cite{Bhattacharya_Viesca_2019,Saez_Lecampion_2022} simplifies the analysis of results. The effect of Poisson's ratio on the rupture shape will be quantified in section \ref{non-circular}.

\subsection{Recall on the self-similar solution before shut-in}
\label{self-similar-solution-circular}
The solution for the continuous-injection stage ($t\leq t_{s}$) of the same problem was presented recently by S\'aez \textit{et al}. \cite{Saez_Lecampion_2022}. We briefly summarize some of their results as they provide the starting point for the understanding of the post-injection phase. S\'aez \textit{et al}. showed that fault slip induced by injection at constant volume rate (from an infinitesimal source) is self-similar in a diffusive manner. The rupture radius $R(t)$ evolves simply as $R(t)=\lambda L(t)$, where $L(t)=\sqrt{4\alpha t}$ is the nominal position of the pore-pressure perturbation front, and $\lambda$ is the so-called amplification factor for which an analytical solution as function of the stress-injection parameter $\mathcal{T}$ exists (eq. (21) in \cite{Saez_Lecampion_2022}). The asymptotic behavior of $\lambda$ for the limiting values of $\mathcal{T}$ is particularly insightful. For critically-stressed faults ($\mathcal{T}\ll 1$), the amplification factor turns out to be large ($\lambda\gg 1$) and thus the rupture front outpaces largely the fluid pressure front ($R(t)\gg L(t)$), with $\lambda$ obeying the simple asymptotic relation $\lambda\approx1/\sqrt{2\mathcal{T}}$. On the other hand, for marginally-pressurized faults ($\mathcal{T}\gg 1$), the amplification factor is small ($\lambda\ll 1$), such that the rupture front lags significantly the fluid pressure front ($R(t)\ll L(t)$), with $\lambda\approx\left(1/2\right)\exp\left[\left(2-\gamma-\mathcal{T}\right)/2\right]$, where $\gamma=0.577216...$ is the Euler-Mascheroni’s constant. These simple closed-form expressions provide important insights into the response of aseismic faults during continuous fluid injections. They encapsulate the dependence of the slip front position on the initial state of stress ($\tau_{0}$, $\sigma_{0}^{\prime}$), constant friction coefficient ($f$), injection rate ($Q$), and fluid and fault hydraulic properties ($\eta$, $kw$, $\alpha$), in a compact and convenient way to determine whether aseismic-slip stress transfer ($\lambda>1$) or pore pressure increase ($\lambda<1$) is the dominant mechanism in the possible triggering of seismicity associated with this type of coupled fluid flow and fault slip processes.

\subsection{Transition from crack-like to pulse-like rupture: the locking front}
In section \ref{pore-pressure}, we showed that the shut-in of the injection is characterized by the emergence of the so-called pore-pressure back front, which corresponds to the instantaneous position at which the pressure has reached its maximum historically. With reference to figure \ref{fig:pp-solution}c, we consequently see that pore pressure is instantaneously increasing at every point on the fault plane located beyond the position of this front: $\partial\Delta p(r,t)/\partial t>0$ for all $r>P(t)$. This results in a moving region where the fault shear strength $\tau_s(r,t)=f\left(\sigma_{0}^{\prime}-\Delta p(r,t)\right)$ is decreasing. Such reduction of fault strength is indeed what uniquely drives the propagation of the rupture after shut-in. On the other hand, in the complementary inner region $r<P(t)$, the opposite holds, and $\partial\Delta p(r,t)/\partial t<0$ (see figure \ref{fig:pp-solution}c). The pore pressure is thus diminishing and the effective normal stress and thus fault shear strength is rising as a result. Furthermore, the enhanced fault shear strength may eventually become locally greater than the resolved shear stress $\tau$, consequently re-locking the slip surface. We observe the development of this re-locking process systematically in all our numerical solutions, starting virtually right after shut-in. Such re-locking process is characterized by a locking front $\mathcal{B}(t)$ propagating outwardly from the injection location. The emergence of this locking front marks the transition from crack-like (before shut-in) to pulse-like (after shut-in) propagation mode, a prominent feature of aseismic slip after shut-in.
\begin{figure}
\centering
  \includegraphics[width=13cm]{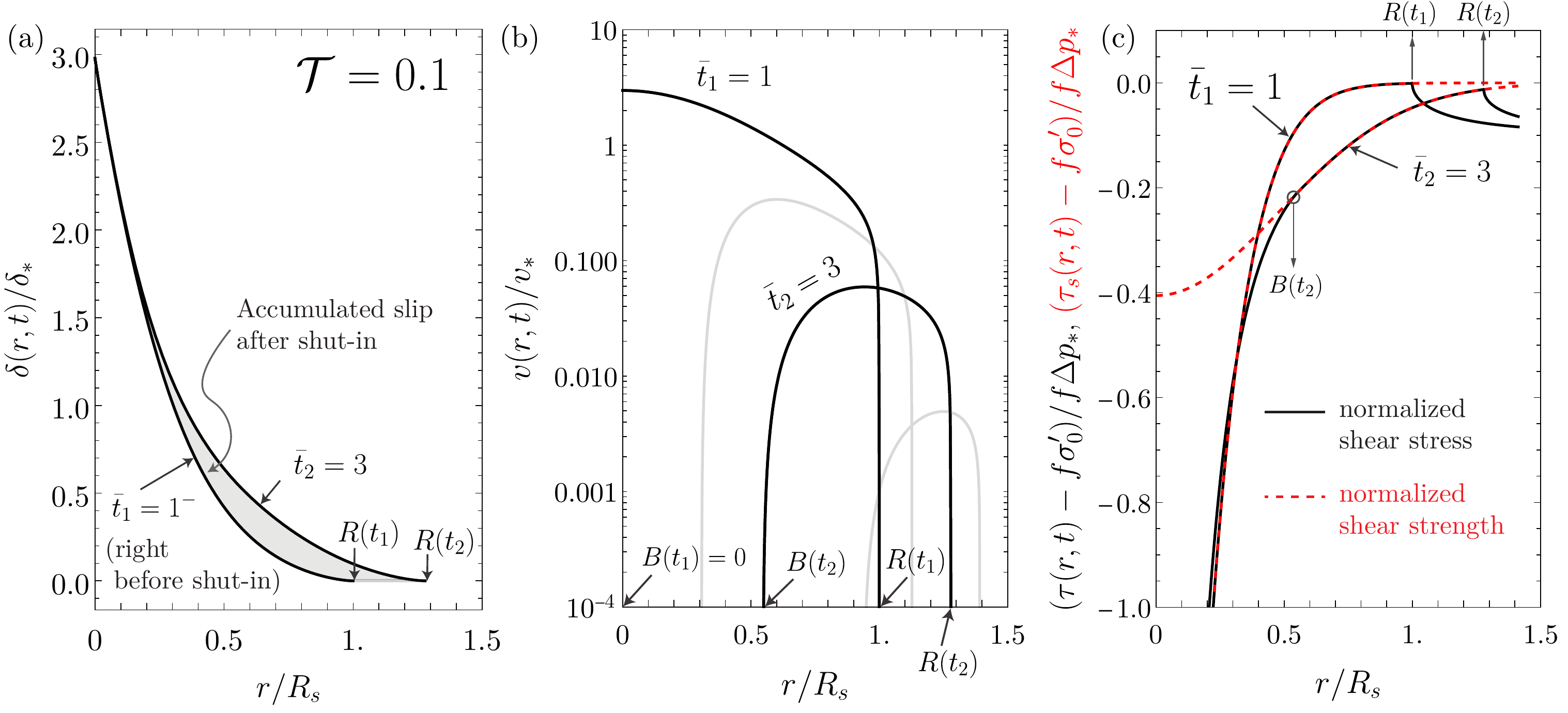}
  \caption{Transition from crack-like to pulse-like rupture on a mildly critically-stressed fault with $\mathcal{T}=0.1$. Normalized spatial profiles of (a) slip $\delta$, (b) slip rate $v$, and (c) shear stress $\tau$ and shear strength $\tau_s$, at two dimensionless times, one right before shut-in ($\bar{t}_1=1$) which is the last moment of crack-like propagation, and the other one at some moment after shut-in ($\bar{t}_2=3$) when pulse-like propagation is taking place. The positions of the locking $B(t)$ and rupture $R(t)$ fronts are highlighted throughout panels (a) to (c). In panel (b), gray curves correspond to aseismic pulses at times $\bar{t}=1.54$ and $6.25$.}
  \label{fig:transition}
\end{figure}

Similarly to the rupture front $\mathcal{R}(t)$, the locking front $\mathcal{B}(t)$ must be circular when $\nu=0$, and is thus fully defined by its radius $B(t)$. Figure \ref{fig:transition} shows the transition from crack-like to pulse-like propagation as seen in the normalized slip $\delta$, slip rate $v$, and shear stress $\tau$ and strength $\tau_s$ spatial distributions, at times right before shut-in and at some time after shut-in. The aseismic pulses can be seen clearly in Figure \ref{fig:transition}b for the slip rate distribution, where the positions of the locking and rupture fronts are highlighted. Note that both the magnitude (peak slip rate) and width of the pulses decrease monotonically with time.  On the other hand, the Mohr-Coulomb inequality given by equation \eqref{eq:MC-criterion} establishes that, within the slipping patch ($B(t)<r<R(t)$), the shear stress must be equal to the fault shear strength, whereas in the re-locked region ($r<B(t)$), the shear stress must be lower than the fault shear strength. Both cases can be clearly seen in figure \ref{fig:transition}c, where the positions of the rupture and locking fronts are also highlighted. Note that the normalized shear strength in figure \ref{fig:transition}c is equal to $(\tau_s-f\sigma_0^\prime)/f\Delta p_*=-\Delta p/\Delta p_*$, which is minus the normalized pore pressure perturbation. In addition, we observe that at some distance away from the locking front and towards the injection point, the shear stress after shut-in is slightly greater than the shear stress right before shut-in. This amplification of fault shear stress after shut-in is due to the amount of slip accumulated through the passage of the slip pulse (shown in figure \ref{fig:transition}a) and the non-local redistribution of stresses associated with it via the quasi-static non-local integral operator of equation \eqref{eq:momentum}. We further elaborate on the amplification of shear stress behind the locking front in section \ref{slip-rate-stress-rate}.

\subsection{Evolution of the rupture, locking, and pore-pressure back fronts: conditions for pulse-like propagation and arrest}
\label{evolution-fronts}
\begin{figure}
\centering
  \includegraphics[width=13cm]{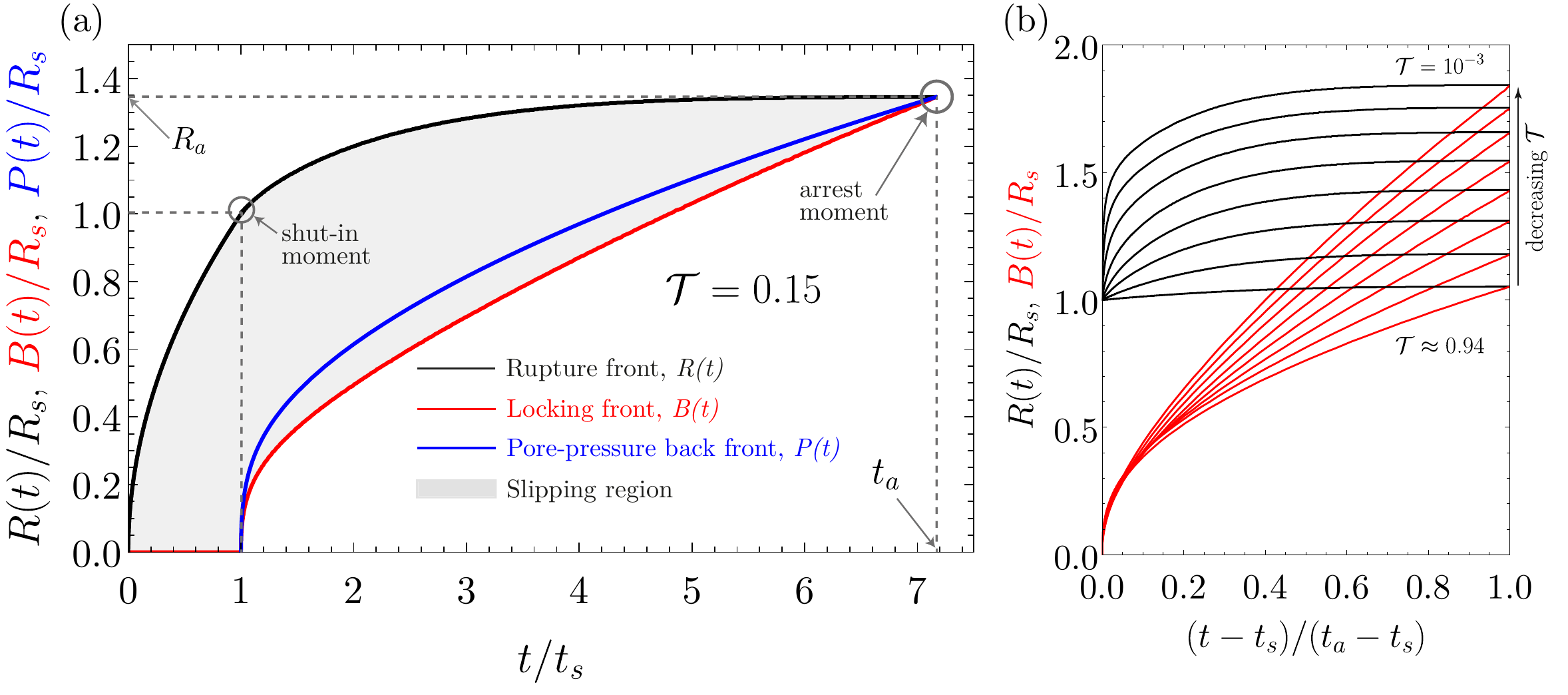}
  \caption{(a) Evolution of the rupture front radius, locking front radius, and pore-pressure back front radius, for an exemplifying case with $\mathcal{T}=0.15$. The front radii are normalized by the rupture radius at the moment of shut-in $R_{s}$ (known analytically from the solution before shut-in, section \ref{self-similar-solution-circular}). Instants of shut-in of the injection $t_{s}$ and arrest of the rupture $t_{a}$ are highlighted, as well as the maximum run-out distance $R_{a}$ (rupture radius at arrest). (b) Evolution of the rupture front radius and locking front radius for many values of $\mathcal{T}$ ranging from $\approx0.94$ to $10^{-3}$. In this plot, time is scaled to be between 0 at shut-in and 1 at arrest.}
  \label{fig:fronts}
\end{figure}

We have identified three relevant fronts that propagate simultaneously upon the stop of the injection: the rupture front of radius $R(t)$, the pore-pressure back front of radius $P(t)$, and the locking front of radius $B(t)$. The temporal evolution of these three fronts is shown in figure \ref{fig:fronts}a for an exemplifying case with $\mathcal{T}=0.15$. Since the only external action that drives the rupture pulses is the further increase of pore pressure after shut-in at distances $r>P(t)$, the pore-pressure back front must be located behind the rupture front in order to guarantee the sustained propagation of the rupture, in other words $P(t)<R(t)$. Similarly, the propagation of the locking front is driven by the existence of a region within the pulse where the pore pressure decreases (and thus the shear strength increases to effectively re-lock the fault), such that $P(t)>B(t)$. Hence, the following inequalities must be satisfied during the propagation of the pulses in the post-injection stage, $B(t)<P(t)<R(t)$, which are verified in all our numerical solutions, such as the one displayed in figure \ref{fig:fronts}a for example.

On the other hand, since the Coulomb's shear strength on the fault plane is always bounded, the shear stress developed at the front of the rupture pulse must be bounded as well, to effectively allow both quantities to be equal. In other words, the slipping patch must propagate with no stress singularity at the "fracture" front of such a cohesive-like crack \cite{Barenblatt_1962}. Hence, both stress intensity factors in mode II and mode III must be equal to zero at $r=R(t)$. This fracture-mechanics-like propagation condition was derived in Appendix A of S\'aez \textit{et al}. \cite{Saez_Lecampion_2022} for any axisymmetric circular shear rupture, and is still valid for pulse-like (annular) ruptures as long as the "fracture" is understood not as the current slipping patch, but rather as the region that has ever experienced any amount of slip since the start of the injection (i.e., for all $r<R(t)$). The condition reads as \cite{Saez_Lecampion_2022}
\begin{equation}\label{eq:fm-cond-Rfront}
    \int_{0}^{R(t)}\frac{\Delta\tau\left(r,t\right)}{\sqrt{R(t)^{2}-r^{2}}}r\textrm{d}r=0\iff \int_{0}^{B(t)}\frac{\tau_{0}-\tau^{*}\left(r,t\right)}{\sqrt{R(t)^{2}-r^{2}}}r\textrm{d}r+\int_{B(t)}^{R(t)}\frac{\tau_{0}-f\left(\sigma_{0}^{\prime}-\Delta p(r,t)\right)}{\sqrt{R(t)^{2}-r^{2}}}r\textrm{d}r=0.
\end{equation}
Note that $\Delta\tau\left(r,t\right)=\tau_{0}-\tau\left(r,t\right)$, with $\tau_{0}$ the initial shear stress and $\tau\left(r,t\right)$ the instantaneous shear stress distribution. In the right-hand side of equation \eqref{eq:fm-cond-Rfront}, we have split the integral into two parts recognizing that within the slipping patch ($B(t)\leq r\leq R(t)$), the shear stress must be equal to the fault shear strength:  $\tau\left(r,t\right)=f\left(\sigma_{0}^{\prime}-\Delta p(r,t)\right)$. In the re-locked region ($r<B(t)$), $\tau^{*}\left(r,t\right)$ corresponds to the instantaneous shear stress distribution which depends on the instantaneous slip distribution over the entire fault plane through the quasi-static non-local integral operator of equation \eqref{eq:momentum}.

The second propagation condition for the post-injection aseismic pulses comes from the analysis of Garagash \cite{Garagash_2012} for seismic and aseismic pulses driven by thermal pressurization. As we will discuss in detail in section \ref{slip-rate-stress-rate}, equation \eqref{eq:momentum-rate} shows that our aseismic pulses are akin to an annular "crack" of inner radius $B(t)$ and outer radius $R(t)$, as long as the dislocation density is replaced by the slip rate gradient $\partial v/\partial r$, and the change of shear stress by the shear stress rate $\partial \tau /\partial t$. The near-tip asymptotic behavior of classical cracks \cite{Rice_1968} is thus valid but in terms of $\partial \tau /\partial t$ and $v$. Having this result in mind, we can now invoke Garagash's healing condition \cite{Garagash_2012}, which states that the shear stress rate $\partial \tau / \partial t$ must be non-singular at the locking front to effectively allow the fault to re-lock. This is because within the slipping patch and, particularly, at the locking front, the shear stress rate must equal the shear strength rate (the plastic consistency condition, see for example, Appendix A in \cite{Saez_Lecampion_2022}), in our case, $\partial \tau / \partial t =-f\partial \Delta p/ \partial t$. Since $\partial \Delta p/ \partial t$ is bounded all over the fault plane, it follows that $\partial \tau / \partial t$ cannot be singular at the locking front. 

Garagash obtained an integral equation similar to \eqref{eq:fm-cond-Rfront} based on the non-singularity of shear stress rate at the locking front (equation (19) in \cite{Garagash_2012}). Such integral equation is valid under the assumptions of his study in 2D elasticity for a solitary steady pulse traveling in anti-plane (III) or in-plane (II) mode of sliding. Indeed, if our pulse that is non-steady would be solved in 2D elasticity, a similar expression can be derived for the two symmetric pulses (traveling in opposite directions) that would emerge in that case, from known stress intensity factors (SIF) formulae \cite{Tada_Paris_Irwin_2000}. To the best of our knowledge, the SIF formulae for an annular crack under axisymmetric loading are unknown (numerical procedures have been proposed to calculate them in mode I, \cite{Clements_Ang_1988}). Nevertheless, one can still formalize this second propagation condition based on the near-tip asymptotic behavior of classical cracks as \cite{Rice_1968,Barenblatt_1962} 
\begin{equation}\label{eq:fm-cond-Bfront}
    K_{\dot{\tau}}=0,\;\text{with}\;\;\frac{\partial\tau(r,t)}{\partial t}\sim\frac{K_{\dot{\tau}}}{\sqrt{2\pi \left(B(t)-r\right)}}\;\text{when}\;\;r \to B(t)^{-}.
\end{equation}
$K_{\dot{\tau}}$ is the so-called "stress-rate intensity factor" \cite{Garagash_2012} with units of $\left(\textrm{pressure/time}\right)\cdot \textrm{length}^{\textrm{1/2}}$. It quantifies the intensity of the leading-order (singular) term of $\partial \tau/\partial t$ near the locking front. Note that $K_{\dot{\tau}}$ may be also written in terms of the mode II, $K_{2,\dot{\tau}}$, and mode III, $K_{3,\dot{\tau}}$, stress-rate intensity factors, as $K_{\dot{\tau}}^2=K_{2,\dot{\tau}}^2+K_{3,\dot{\tau}}^2$. Indeed, since the two components of the shear stress rate on the fault plane must be non-singular along the locking front, equation \eqref{eq:fm-cond-Bfront} can be stated in the stronger form, $K_{2,\dot{\tau}}=K_{3,\dot{\tau}}=0$. Note also that the stress-rate intensity factor is not the same than the stress intensity factor rate, at least along the locking front. For the latter, there is no stress intensity factor whatsoever, since there is no fracture tip and thus no potential singular term for $\tau$ when $r \to B(t)^{-}$. Hence, although it could have been tempting to write the healing condition \eqref{eq:fm-cond-Bfront} as $\text{d}\mathcal{G}/\text{d}t=0$, where $\mathcal{G}$ is the energy release rate, such statement is not true. 

Equations \eqref{eq:fm-cond-Rfront} and \eqref{eq:fm-cond-Bfront} are, therefore, the necessary conditions for the propagation of post-injection aseismic slip as an annular pulse. Together, these two equations describe the motion of the rupture front $R(t)$ and locking front $B(t)$ that are at quasi-static equilibrium with a given $\Delta \tau (r,t)$. Yet as already mentioned, we cannot solve these two equations without solving the entire moving boundary value problem. However, they provide additional insights into the mechanics of the problem and, as we will see later, will define important characteristics of the slip rate and shear stress rate distributions along the fault plane.

We now seek for the conditions that characterize the final arrest of the rupture pulses. Figure \ref{fig:fronts}a shows that as expected at the moment of arrest, the locking front $B(t)$ catches up the rupture front $R(t)$. Since in that instant the rupture pulse effectively vanishes, equation \eqref{eq:fm-cond-Bfront} has no relevance. However, the "fracture" (slipped surface) is still present and, to be at mechanical equilibrium,  equation \eqref{eq:fm-cond-Rfront} must be satisfied. Since the second term of the right-hand side of \eqref{eq:fm-cond-Rfront} approaches zero at the time of arrest, equation \eqref{eq:fm-cond-Rfront} takes now the simpler form
\begin{equation}\label{eq:arrest-cond-1}
    \int_{0}^{R_{a}}\frac{\tau_{0}-\tau^{*}\left(r,t_{a}\right)}{\sqrt{R_{a}^{2}-r^{2}}}r\textrm{d}r=0,
\end{equation}
where $t_{a}$ is the time of arrest and $R_{a}=R(t_{a})$ is the corresponding rupture radius at the arrest time (the maximum run-out distance). In the previous equation, $R_{a}$ may be equivalently replaced by $B_{a}=B(t_{a})$. Furthermore, since the rupture after shut-in is driven uniquely by the further increase of pore pressure that occurs in the region $r>P(t)$, the rupture arrest will be also characterized by the moment at which the pore-pressure back front catches up the rupture front (see figure \ref{fig:fronts}a). Therefore, the following conditions must be also met exactly at the time of arrest, 
\begin{equation}\label{eq:R-B-P-equal}
    R(t_{a})=B(t_{a})=P(t_{a}),
\end{equation}
 and more importantly, by combining the previous equation with equation \eqref{eq:P-front}, we obtain an analytic relation between the radius $R_{a}$ and time $t_{a}$ of arrest,
 \begin{equation}\label{eq:arrest-cond-2}
     R_{a}=\sqrt{4\alpha t_{a}}\left(\left(t_{a}/t_{s}-1\right)\ln\left(\frac{t_{a}}{t_{a}-t_{s}}\right)\right)^{1/2},
 \end{equation}
 which in the large-time limit ($t_{a}\gg t_{s}$) becomes simply $R_{a}\approx\sqrt{4\alpha t_{a}}$. Equations \eqref{eq:arrest-cond-1} and \eqref{eq:arrest-cond-2} provide a complete system of equations to solve for the time at arrest $t_{a}$ and the maximum run-out distance $R_{a}$. However, the shear stress profile left by the arrested rupture $\tau^{*}\left(r,t_{a}\right)$ in equation \eqref{eq:arrest-cond-1} is unknown analytically. Hence, we must determine at least one of these two relevant quantities numerically.

\subsection{Arrest time and maximum run-out distance}
\begin{figure}
\centering
  \includegraphics[width=13cm]{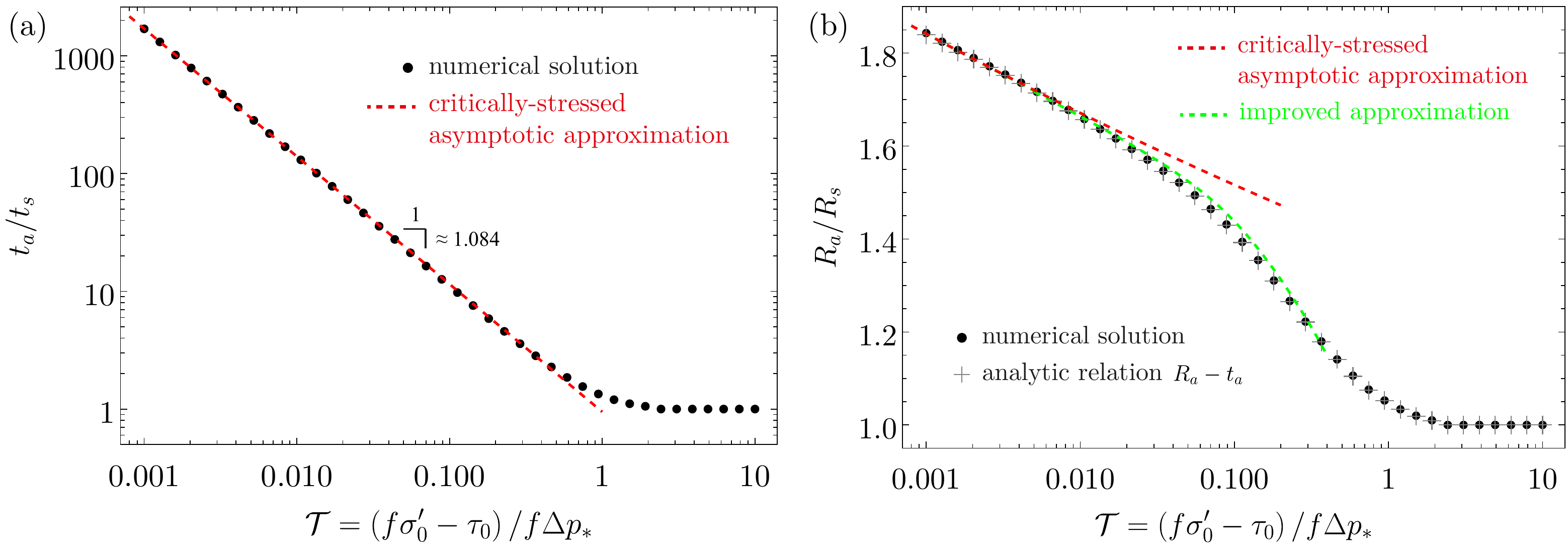}
  \caption{(a) Normalized arrest time $t_{a}/t_{s}$ and (b) normalized maximum run-out distance $R_{a}/R_{s}$, as a function of the stress-injection parameter $\mathcal{T}$. In (a), the red dashed line represents the asymptotic behavior in the critically-stressed regime ($\mathcal{T}\ll1$), in the form of a numerically-derived power law, equation \eqref{eq:ta-power-law}. In (b), the gray crosses correspond to the evaluation of $R_{a}$ via the analytic relation, equation \eqref{eq:arrest-cond-2}, using the numerical solution of $t_{a}$. The red dashed curve corresponds to the critically-stressed asymptotic approximation, equation \eqref{eq:Ra-asymptote}, whereas the green dashed curve is an improved approximation obtained by substituting equation \eqref{eq:ta-power-law} directly into \eqref{eq:arrest-cond-2}.}
  \label{fig:ta-Ra-T}
\end{figure}
We now determine numerically the normalized arrest time $t_{a}/t_{s}$ and normalized maximum run-out distance $R_{a}/R_{s}$ by spanning the relevant parameter space with 40 values of $\mathcal{T}$ ranging from $10^{-3}$ to $10$, equally spaced in logarithmic scale. The results are presented in figure \ref{fig:ta-Ra-T}. Marginally-pressurized faults ($\mathcal{T}\gg1$) produce slip pulses that arrest almost immediately after the stop of the injection and practically do not grow further than the size of the ruptures at the moment of shut-in ($R_s$), whereas critically-stressed faults ($T\ll1$) host events that can last up to $\approx10^{3}$ times the injection time $t_s$ and grow approximately up to 1.8 times the rupture radius at the shut-in time (for the smallest value of $\mathcal{T}$ considered). Critically-stressed faults are, therefore, the asymptotic regime of more practical interest. We thus focus on determining its behavior precisely. 

For the normalized arrest time, figure \ref{fig:ta-Ra-T}a shows a clear power law between $t_{a}/t_{s}$ and $\mathcal{T}$. Using a linear least square regression, we obtain
\begin{equation}\label{eq:ta-power-law}
    \frac{t_{a}}{t_{s}}=a\mathcal{T}^{-b},
\end{equation}
with $a=0.946876$ and $b=1.084361$. In solving the least-squares problem, we have considered data points satisfying $T\lesssim0.01$ ($\ll1$) and obtained a coefficient of determination $R^{2}$ equal to $0.999993$.

We can now use the analytic relation between $R_{a}$ and $t_{a}$, equation \eqref{eq:arrest-cond-2}, to construct a critically-stressed asymptotic approximation for the normalized maximum run-out distance using the power-law relation \eqref{eq:ta-power-law}. In particular, at large arrest times ($t_{a}\gg t_{s}$), \eqref{eq:arrest-cond-2} is approximately equal to $R_{a}/R_{s}\approx\left(1/\lambda\right)\sqrt{t_{a}/t_{s}}$. Substituting $\lambda= 1/\sqrt{2\mathcal{T}}$ (see section \ref{self-similar-solution-circular}) and \eqref{eq:ta-power-law} into the previous expression, leads to the sought critically-stressed approximation for the maximum run-out distance,
\begin{equation}\label{eq:Ra-asymptote}
    \frac{R_{a}}{R_{s}}\approx\left(\frac{2a}{\mathcal{T}^{b-1}}\right)^{1/2},\;\textrm{or}\;\;R_{a}\approx\left(\frac{4a\alpha t_{s}}{\mathcal{T}^{b}}\right)^{1/2}.
\end{equation}
In addition, we can substitute equation \eqref{eq:ta-power-law} directly into \eqref{eq:arrest-cond-2} to provide an improved asymptotic approximation for the critically-stressed regime that is approximately valid over a broader range of values of $\mathcal{T}$ (green curve in figure \ref{fig:ta-Ra-T}b). Note that in figure \ref{fig:ta-Ra-T}b, we have also plotted the values of $R_{a}$ not from the numerical solution, but rather by evaluating the analytic relation \eqref{eq:arrest-cond-2} (gray crosses) given the numerical solution of $t_{a}$. The latter is just to illustrate the exactness of the relation between the radius of arrest and the time of arrest \eqref{eq:arrest-cond-2}.

\subsection{Characteristics of slip rate and shear stress rate}
\label{slip-rate-stress-rate}
\begin{figure}
\centering
  \includegraphics[width=12.85cm]{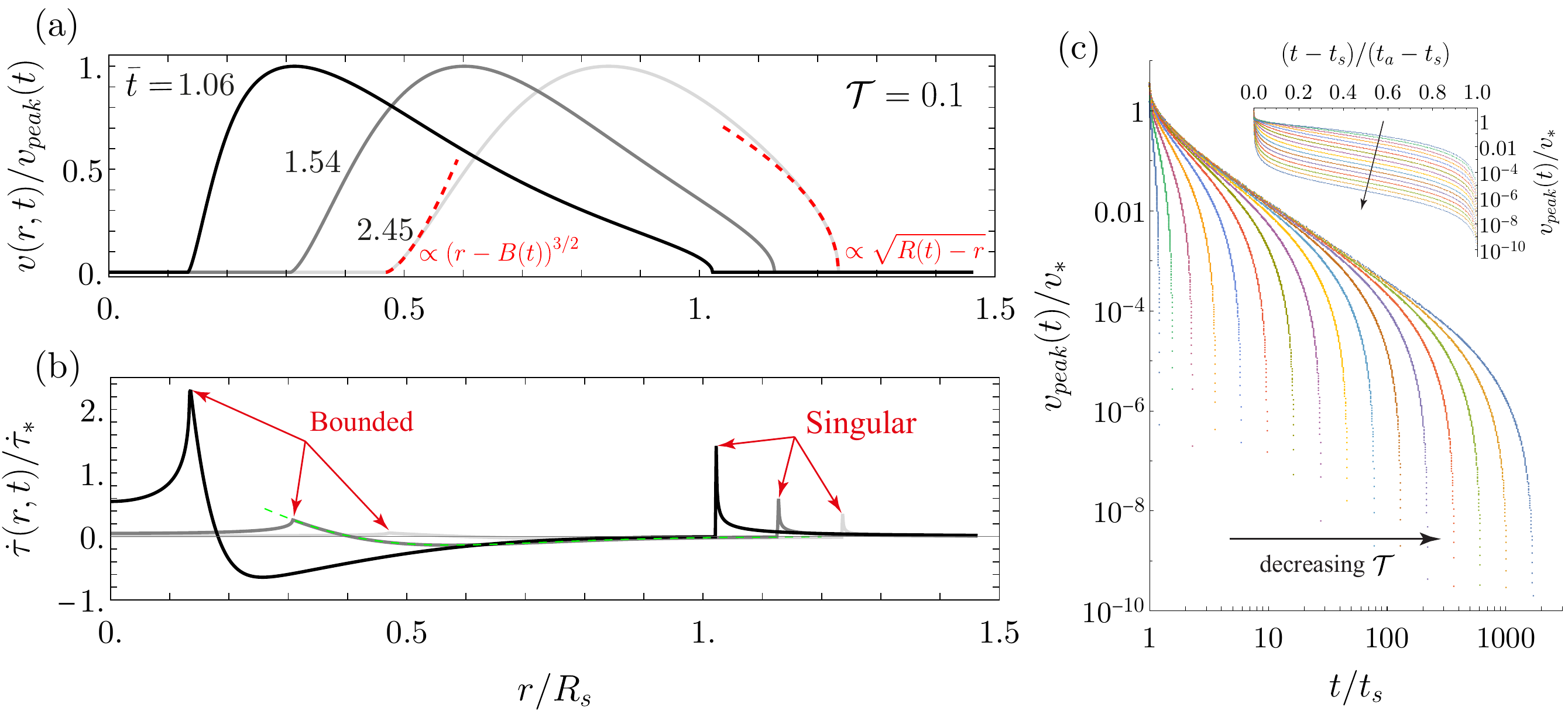}
  \caption{(a) Slip rate distribution $v$ normalized by the instantaneous peak slip rate $v_{peak}(t)$ at three dimensionless times for a mildly critically-stressed fault with $\mathcal{T}=0.1$. Red dashed curves correspond to the tip asymptotics, equation \eqref{eq:tip-asymptotics-slip-rate}. (b) Shear stress rate distributions normalized by the stress-rate scale $\dot{\tau}_*=f\Delta p_*/t_s$, for the same three dimensionless times in (a). The singular and non-singular stress rate fields at the rupture and locking fronts, respectively, are highlighted in red. The green dashed curve corresponds to the normalized rate of shear strength at $\bar{t}=1.54$. (c) Evolution of the normalized peak slip rate $v_{peak}$ in time for many values of $\mathcal{T}$ ranging from $\approx 0.75$ to $10^{-3}$ (from left to right). In the inset, same results but time is scaled to be between $0$ at shut-in and $1$ at arrest. The black arrows indicate the direction of decreasing $\mathcal{T}$ (pointing towards the critically-stressed limit).}
  \label{fig:slip-rate-characteristics}
\end{figure}

We now examine in detail the slip-rate characteristics of the aseismic pulses (figure \ref{fig:slip-rate-characteristics}). Let us consider the quasi-static elastic equilibrium that relates the fault shear stress $\tau$ and slip distribution $\delta$ for an axisymmetric circular shear rupture, in the compact form given by Bhattacharya and Viesca \cite{Bhattacharya_Viesca_2019} (derived from \cite{Salamon_Dundurs_1977}, and also reproduced in Appendix C of S\'aez \textit{et al}. \cite{Saez_Lecampion_2022}). We differentiate that integral equation with respect to time (using Leibniz’s integral rule, applying the condition of no shear stress singularity at the rupture front $\partial\delta\left(R(t),t\right)/\partial r=0$, and the fact that $\partial v/\partial r=0$ within the re-locked region $0<r<B(t)$) to arrive to the following relation\footnote{albeit with a minus one factor due to the difference between the geomechanics convention of positive stresses in compression and the classical solid mechanics convention of positive stresses in tension.},
\begin{equation}\label{eq:momentum-rate}
    \frac{\partial\tau\left(r,t\right)}{\partial t}=\frac{\mu}{2\pi}\int_{B(t)}^{R(t)}\frac{\partial v(\xi,t)}{\partial\xi}\left(\frac{K\left[k(r/\xi)\right]}{\xi+r}+\frac{E\left[k(r/\xi)\right]}{\xi-r}\right)\mathrm{d}\xi,
\end{equation}
where $K\left[ \cdot \right]$ and $E\left[ \cdot \right]$ are the complete elliptic integrals of the first and second kind, respectively, and $k(x)=2\sqrt{x}/\left(1+x\right)$. Equation \eqref{eq:momentum-rate} is akin to the elastic equilibrium of an annular crack of inner radius $B(t)$ and outer radius $R(t)$, but in terms of shear stress rate $\partial \tau /\partial t$ and slip rate gradient $\partial v/\partial r$, instead of shear stress $\tau$ and slip gradient $\partial \delta / \partial r$. The asymptotic behavior of the slip rate distribution near the rupture and locking fronts thus follows the tip asymptotics for displacement discontinuity of classical cracks \cite{Rice_1968,Barenblatt_1962},
\begin{equation}\label{eq:tip-asymptotics-slip-rate}
    v(r,t) \propto \sqrt{R(t)-r}\;\;\text{when}\;\;r \to R(t)^{-}\;\text{, and}\;\; v(r,t) \propto \left(r-B(t)\right)^{3/2}\;\;\text{when}\;\;r \to B(t)^{+}.
\end{equation}
Note that the rupture-front asymptotic features a stress-rate singularity in the classic form $\partial \tau /\partial t\sim K_{\dot{\tau},R}/\sqrt{2\pi \left(r-R(t)\right)}$ when $r \to R(t)^{+}$, where $K_{\dot{\tau},R}$ is the "stress-rate intensity factor". The subscript $R$ is introduced to distinguish $K_{\dot{\tau},R}$ from the stress-rate intensity factor of the locking front $K_{\dot{\tau}}$ already introduced in equation \eqref{eq:fm-cond-Bfront}. Note that the singularity may be written in a stronger form for the two components of shear stress rate on the fault plane (the mode II and mode III directions), as already discussed for the locking front in section \ref{evolution-fronts}. On the other hand, the healing condition \eqref{eq:fm-cond-Bfront} has been already taken into account in equation \eqref{eq:tip-asymptotics-slip-rate} for the locking front, such that the leading-order term of the slip rate is now the next term of the near-tip asymptotic expansion of classical cracks $\dot{\delta}\propto s^{3/2}$\cite{Rice_1968,Barenblatt_1962}, where $s$ is the distance from the tip towards the slipping patch. In this way, the finiteness of the shear stress rate at the locking front is achieved, whereas nothing prevents the shear stress rate to be infinite at the rupture front. It is indeed the shear stress $\tau$ the one quantity that must be bounded at the rupture front, according to the condition \eqref{eq:fm-cond-Rfront}.

The foregoing tip asymptotics are highlighted in red in figures \ref{fig:slip-rate-characteristics}a-b, which show typical slip pulses and their corresponding shear stress rate distributions along the fault plane at three dimensionless times. In figure \ref{fig:slip-rate-characteristics}b, the green dashed curve corresponds to the normalized rate of shear strength at $\bar{t}=1.54$ that must be equal to the normalized shear stress rate within the slipping patch according to the  consistency condition mentioned in section \ref{evolution-fronts}. Such equality can be clearly visualized in this figure. Note that the slip rate in figure \ref{fig:slip-rate-characteristics}a is normalized by the instantaneous maximum slip rate $v_{peak}(t)$, that decays fast in time. The evolution of the maximum slip rate is indeed displayed in figure \ref{fig:slip-rate-characteristics}c for relevant values of $\mathcal{T}$ ranging from $10^{-3}$ to $\approx 0.75$ (greater values than $0.75$ correspond to ruptures that arrest almost immediately after shut-in). Owing to the relatively fast decay of the maximum slip rate after shut-in, the peak shear stress rate at both the locking and rupture fronts is also observed to decay fast, as depicted in figure \ref{fig:slip-rate-characteristics}b.

\section{Pulse-like non-circular ruptures}
\label{non-circular}
In this section, we examine the effect of a Poisson's ratio different than zero on the propagation and ultimate arrest of the post-injection aseismic pulses. For this purpose, we consider a fixed value of $\nu=1/4$ (a so-called Poisson's solid) and solve the governing equations via our fully 3D numerical solver.

\subsection{Recall on the self-similar solution before shut-in}
\label{self-similar-solution-non-circular}
S\'aez \textit{et al}. \cite{Saez_Lecampion_2022} also solved the continuous-injection problem for the case of $\nu\neq0$. They showed that the rupture front is well-approximated by an elliptical shape that becomes more elongated for increasing values of $\nu$ and decreasing values of $\mathcal{T}$: the more critically stressed the fault is, the more elongated the rupture becomes. The aspect ratio of the quasi-elliptical rupture fronts is upper bounded by $1/(1-\nu)$ in the critically-stressed limit ($\mathcal{T}\ll 1$), and lower bounded by $(3-\nu)/(3-2\nu)$ in the marginally-pressurized limit ($\mathcal{T}\gg 1$). Furthermore, the rupture area $A_{r}(t)$ evolves simply as $A_{r}(t)=4\pi\alpha\lambda^{2}t$ (linear with time), and interestingly it does not depend on the Poisson's ratio $\nu$. The previous numerical observations led to closed-form approximate expressions for the entire evolution of the quasi-elliptical rupture fronts in the terms of the semi-major $a(t)$ and semi-minor $b(t)$ axes of an ellipse, as $a(t)=R(t)/\sqrt{1-\nu}$ and $b(t)=R(t)\sqrt{1-\nu}$ in the critically-stressed regime, and $a(t)=R(t)\sqrt{3-\nu}/\sqrt{3-2\nu}$ and $b(t)=R(t)\sqrt{3-2\nu}/\sqrt{3-\nu}$ in the marginally-pressurized regime, with $R(t)$ the rupture radius of a circular rupture for the same value of $\mathcal{T}$, which is known analytically (see section \ref{self-similar-solution-circular}).

\subsection{Effect of $\nu$ on the propagation of the aseismic pulses}

\begin{figure}
\centering
  \includegraphics[width=12.9cm]{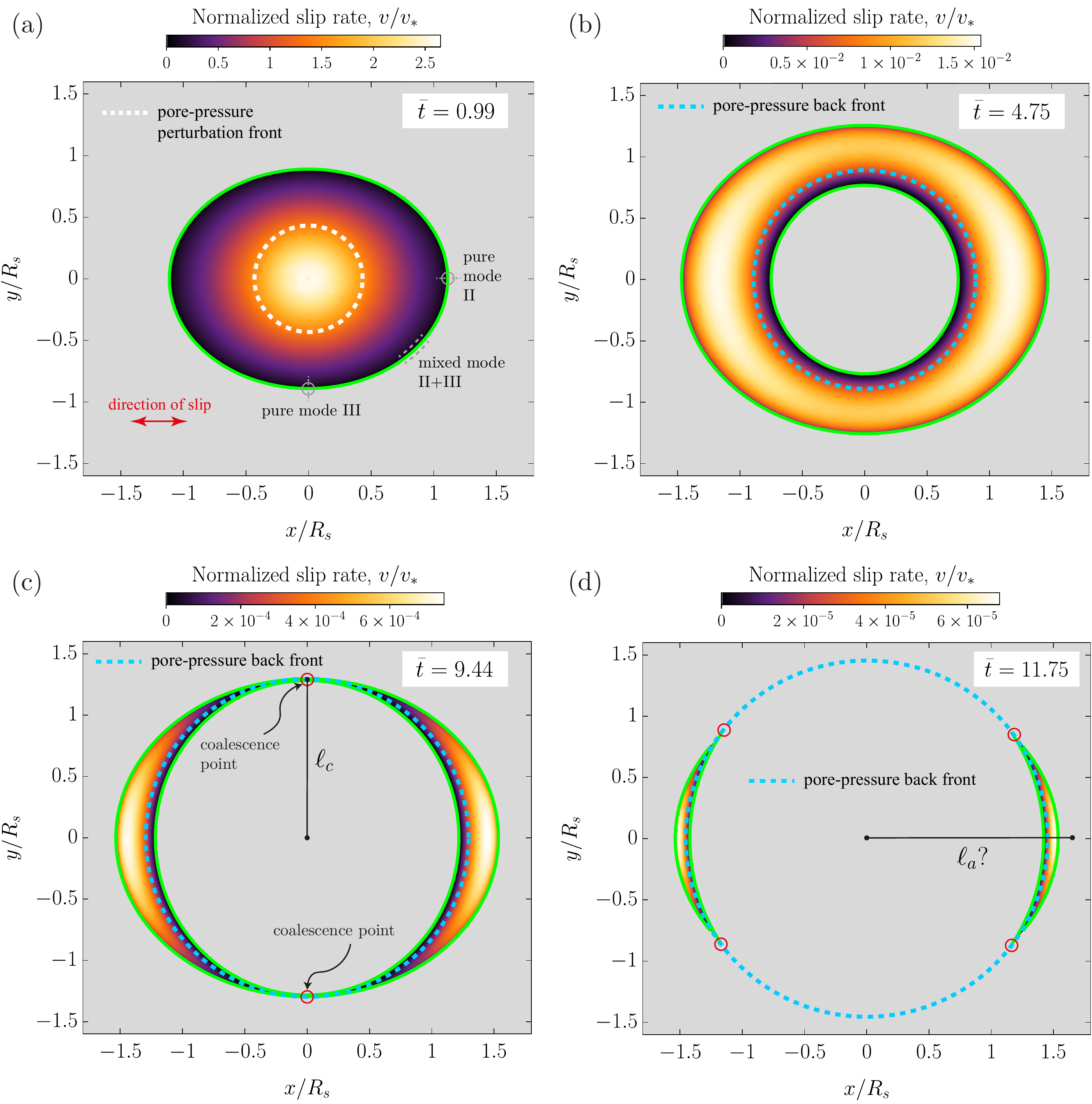}
  \caption{Snapshots of normalized slip rate distribution for a non-circular rupture with $\mathcal{T}=0.1$ and $\nu=0.25$. The spatial coordinates $x$ and $y$ are scaled by the rupture length scale $R_s$ (rupture radius at the moment of shut-in for the same $\mathcal{T}$ but $\nu=0$), whereas time $t$ is normalized by the injection duration $t_s$ as usual. (a) Crack-like propagation right before shut-in ($\bar{t}=0.99$). Direction of slip and mode II and mode III directions of sliding are highlighted. (b) Pulse-like propagation as an annular pulse after shut-in ($\bar{t}=4.75$). (c) Snapshot at the moment in which the locking and rupture fronts are about to coalesce ($\bar{t}\approx9.44$). The maximum run-out distance of the rupture in the mode III direction $\ell_c$ is defined. (d) Pulse-like propagation after coalescence of the fronts as two "moon-shaped" pulses traveling in opposite directions ($\bar{t}=11.75$). The maximum run-out distance of the rupture $\ell_a$ that will be eventually reached at some later time is depicted. The positions at which the locking, rupture, and pore-pressure back fronts intersect each other are indicated with red circles.}
  \label{fig:snapshots}
\end{figure}

We summarize the main characteristics of the propagation of pulse-like non-circular ruptures in figure \ref{fig:snapshots}. This figure is composed by four snapshots of the normalized slip rate distribution for a mildly critically-stressed fault with $\mathcal{T}=0.1$. Each snapshot represents a remarkably different propagation phase. The first one, figure \ref{fig:snapshots}a, corresponds to a moment right before stopping the injection, specifically at $\bar{t}=0.99$, where $\bar{t}=t/t_s$. At this time, the rupture is still propagating in crack-like mode, and for this value of $\mathcal{T}$, the quasi-elliptical rupture front outpaces the pore-pressure perturbation front. Note that the spatial coordinates in figure \ref{fig:snapshots} are normalized by the rupture lengthscale $R_s=\lambda \sqrt{4\alpha t_s}$ (radius of a circular rupture with the same value of $\mathcal{T}$, see section \ref{self-similar-solution-circular}). The second snapshot, figure \ref{fig:snapshots}b, displays the rupture after shut-in at approximately 4.75 times the injection duration. In this figure, we can already observe the locking front and thus the propagation of the rupture as an annular pulse. Interestingly, although the rupture front is clearly elongated along the mode II direction of sliding (same as it is before shut-in), the locking front seems to be slightly elongated along the mode III direction instead.

Figures \ref{fig:snapshots}c and \ref{fig:snapshots}d show the most remarkable effects of considering a Poisson's ratio different than zero. Figure \ref{fig:snapshots}c corresponds to the instant at which the locking front is about to coalesce with the rupture front along the mode III direction of sliding. Note that at this time of coalescence $t_{c}$, approximately equal to 9.44 times the injection duration in this case, the pore-pressure back front is also intersecting both the rupture and locking fronts. Let us denote the maximum run-out distance of the rupture in the mode III direction as $\ell_c$ (see figure \ref{fig:snapshots}c for its definition). The previous observation, that was somewhat expected from the analysis of circular ruptures (see equation \eqref{eq:R-B-P-equal}), means that there is a unique relation between $\ell_c$ and $t_c$, in the form $\ell_c=P(t_c)$, where $P(t)$ is the instantaneous radius of the pore-pressure back front known analytically from equation \eqref{eq:P-front}. Moreover, after the coalescence of the fronts, figure \ref{fig:snapshots}d displays that the original annular pulse rupture is split into two symmetric "moon-shaped" pulses that travel in opposite directions. The moving positions at which the locking and rupture fronts are intersecting each other are shown with red circles in figure \ref{fig:snapshots}d. These positions are again such that the pore-pressure back front is located exactly at the same place. The condition \eqref{eq:R-B-P-equal} for circular ruptures at the time of arrest is thus still valid for non-circular ruptures but now at any time $t_c\leq t\leq t_a$ during this last phase of propagation in which the annular pulse has split. Finally, the instantaneous rupture area of each moon-shaped pulse decreases continuously with time and will eventually collapse into a point located along the \textit{x} axis at some distance $\ell_a$ as depicted in figure \ref{fig:snapshots}d. $\ell_a$ corresponds to the maximum run-out distance of the entire rupture and, similarly to the case of circular ruptures, equation \eqref{eq:arrest-cond-2} will now take the form $\ell_a=P(t_a)$, where $t_a$ is the time of arrest.

\subsection{Effect of $\nu$ on the arrest time and maximum run-out distance}
\begin{figure}
    \centering
    \includegraphics[width=12.8cm]{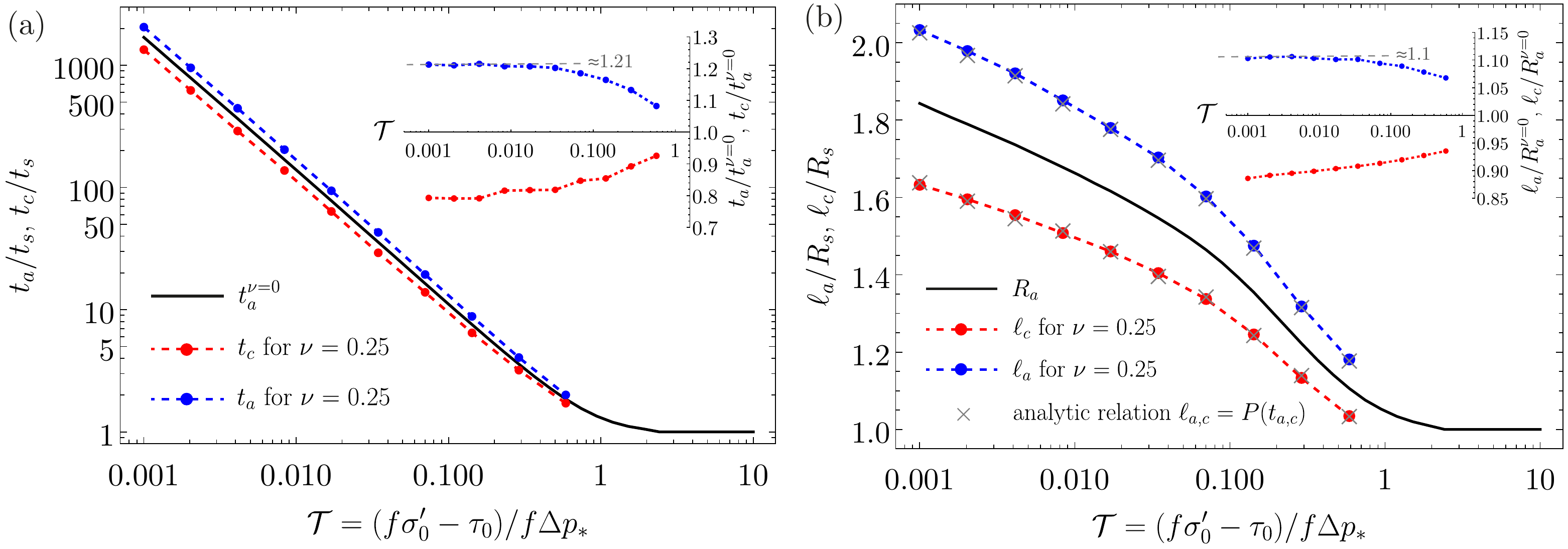}
    \caption{(a) Normalized arrest time $t_a$ and coalescence time $t_c$ for $\nu=0.25$ as a function of the stress-injection parameter $\mathcal{T}$ for values ranging from $10^{-3}$ to $\approx 0.59$. The inset shows the same results but normalized by the arrest time of circular ruptures $t_a^{\nu=0}$. (b) Same as (a) but for the normalized maximum run-out distance of the rupture in the mode II direction of sliding $\ell_a$ and in the mode III direction of sliding $\ell_c$. The inset shows the same results of the main plot but normalized by the maximum run-out distance (or arrest radius) of circular ruptures $R_a$. Black curves represent the circular rupture solution ($\nu=0$) in both panels. Gray crosses in (b) come from evaluating $t_a$ and $t_c$ of panel (a) in equation \eqref{eq:P-front} for the position of the pore-pressure back front.}
    \label{fig:Ra-ta-T-non-circular}
\end{figure}
We calculate the time of arrest $t_a$ and maximum run-out distance of the rupture $\ell_a$, as well as the time of coalescence $t_c$ and maximum run-out distance in the mode III direction $\ell_c$, for ten values of the stress-injection parameter $\mathcal{T}$ ranging from $10^{-3}$ to $\approx 0.59$. This range allows us to span the most relevant part of the parameter space associated with critically-stressed faults. Figure \ref{fig:Ra-ta-T-non-circular}a summarizes the results for the normalized arrest time $t_a/t_s$ and normalized time of coalescence $t_c/t_s$, including the results for the arrest time of circular ruptures $t_a^{\nu=0}$. We observe that the coalescence of the fronts occurs at earlier times than the arrest of circular ruptures. Moreover, a Poisson's ratio $\nu\neq0$ has the clear effect of delaying the arrest of the aseismic pulses with regard to the circular case. To better quantify this delayed arrest, we plot the same results in the inset but normalizing the arrest and coalescence times by the arrest time of circular ruptures. It can be seen that the more critically-stressed the fault is, the longer it takes for non-circular ruptures to arrest compared to circular ones. Furthermore, the ratio $t_a/t_a^{\nu=0}$ seems to approach an asymptotic value of $\approx1.21$ in the critically-stressed limit (when $\mathcal{T} \rightarrow 0$).

Figure \ref{fig:Ra-ta-T-non-circular}b shows the results for the normalized maximum run-out distance of the rupture $\ell_a/R_s$ and normalized maximum run-out distance in the mode III direction $\ell_c/R_s$. In accordance with the results for $t_c$ and $t_a$, the distance $\ell_c$ is lower than the arrest radius of circular ruptures $R_a$ for the same value of $\mathcal{T}$, whereas the maximum run-out distance of non-circular ruptures $\ell_a$ is greater than the maximum run-out distance of its circular counterpart $R_a$. In addition, the relations $\ell_c=P(t_c)$ and $\ell_a=P(t_a)$ discussed in the previous section are shown to be in good agreement (gray crosses in figure \ref{fig:Ra-ta-T-non-circular}b) with the numerical results up to numerical discretization errors. Moreover, the inset shows the run-out distances $\ell_a$ and $\ell_c$ but normalized by the arrested radius of circular ruptures $R_a$. The maximum run-out distance $\ell_a$ seems to approach an asymptotic value in this case of $\approx1.1 R_a$ in the limit of critically-stressed faults.

\section{Discussion}
\label{discussion}
\subsection{Critically-stressed regime versus marginally-pressurized regime}
\label{critically-stressed-versus-marginally-pressurized}
Figures \ref{fig:ta-Ra-T} and \ref{fig:Ra-ta-T-non-circular} summarize one of the most important results of this article. Marginally-pressurized faults ($\mathcal{T}\gg1$) produce aseismic pulses that arrest almost immediately after shut-in, whereas critically-stressed faults ($\mathcal{T}\ll1$) are predicted to host ruptures that propagate for several orders of magnitude ($\approx10^3$) the injection duration and grow up to approximately twice the size of the ruptures at the moment of shut-in (for the smallest value of $\mathcal{T}$ considered). Critically-stressed faults are therefore the relevant propagation regime that is able to sustain seismicity after shut-in. Under this regime, the long-lived aseismic pulses provide a longer exposure of the surrounding rock mass to continuous stressing due to aseismic slip and, at the same time, their longer run-out distances perturb a larger volume of rock mass up to $\approx2^3=8$ times larger than the volume affected at the moment of shut-in, therefore increasing the likelihood of triggering earthquakes during the post-injection stage compared to the marginally-pressurized case.

To illustrate under what conditions these two regimes can occur, let us consider some characteristic values of a fault undergoing fluid injection. Consider that water is injected into a fault zone at a constant volume rate $Q\sim30$ [l/s], which is typical of hydro-shearing treatments of deep geothermal reservoirs at around 3 to 4 km depth \cite{Ellsworth_Giardini_2019}. Assume also a fluid dynamic viscosity $\eta\sim10^{-3}$ [$\text{Pa}\cdot\text{s}$]. The fault hydraulic transmissivity can vary over several orders of magnitude \cite{Wibberley_Shimamoto_2003}, consider, for instance, a plausible value of $kw\sim10^{-13}$ [$\text{m}^{3}$]. This yields an intensity of the injection $\Delta p_*\sim 24$ [MPa] (equation \eqref{eq:p*}). Let us consider for the initial effective normal stress a characteristic value of $\sigma_{0}^{\prime}\sim60$ [MPa], that is somewhat consistent with a depth of 3-4 km under gravitational loading and hydrostatic conditions. Assuming a friction coefficient $f=0.6$, the initial fault shear strength is $f\sigma_0^{\prime}\sim 36$ [MPa], while the intensity of the strength reduction due to fluid injection is $f\Delta p_*\sim 14$ [MPa]. The amount of initial shear stress $\tau_{0}$ acting on the fault will finally determine the regime of the fault response in this example. Consider first a fault that is close to failure, say $\tau_{0}\sim35$ [MPa] ($1$ [MPa] to failure). This leads to a small value of the stress-injection parameter $\mathcal{T}\approx 0.07$ (equation \eqref{eq:T-parameter}), and therefore to a fault responding in the so-called critically-stressed regime ($\mathcal{T}\ll1$). Before shut-in, the fluid-induced aseismic slip front in this example is predicted to outpace the pore pressure perturbation front by a constant factor $\lambda\approx1/\sqrt{2\mathcal{T}}\approx2.7$ (section \ref{self-similar-solution-circular}), for the case of circular ruptures. Consider that the injection was conducted during a period of $t_s\sim1$ [day]. Then, after shut-in, the post-injection aseismic pulse is expected to propagate for $t_a\approx17$ [days] (equation \eqref{eq:ta-power-law}) and reach a maximum run-out distance of $\approx 1.54$ times the run-out distance at the moment of shut-in (equation \eqref{eq:Ra-asymptote}). To close this dimensional example, let us assume that the fault has a hydraulic diffusivity of, say, $\alpha\sim0.01$ [m$^{2}$/s]. The radius of the rupture at the moment of shut-in is $R_s=\lambda \sqrt{4\alpha t_s}\approx 159$ [m] (see again section \ref{self-similar-solution-circular}), and the maximum run-out distance of post-injection aseismic slip will be $R_a\approx 1.54\cdot159$ [m] $\approx 245$ [m]. If the hydraulic diffusivity were ten times lower, $R_s$ and $R_a$ would decrease of approximately 30\%.

Consider the same example of the previous paragraph, but now with a fault that is further away from failure. Assume, for instance, a lower value of the initial shear stress $\tau_0\sim 1$ [MPa]. The stress-injection parameter for this case becomes $\mathcal{T}\approx 2.5$, well within the marginally pressurized regime. Prior shut-in, the slip front is predicted to lag the pore pressure perturbation front by a factor $\lambda \approx \left(1/2\right)\exp[\left(2-\gamma-\mathcal{T}\right)/2]\approx 0.29$ (section \ref{self-similar-solution-circular}). Moreover, upon the end of the injection, the rupture is expected to arrest almost immediately (see figure \ref{fig:ta-Ra-T}), such that $t_a\approx t_s\approx 1$ [day] and $R_a\approx R_s=\lambda\sqrt{4\alpha t_s}\approx 17$ [m]. In the two previous examples, the end-member regimes were achieved by changing only the initial shear stress acting on the fault or, equivalently, the distance to failure. This highlights the importance of the pre-injection stress state on determining the fault response. However, there is another relevant quantity that may equally change the fault response regime, namely, the intensity of the injection $\Delta p_*$. So far, we have assumed hydraulic parameters resulting in $\Delta p_*\sim 24$ [MPa]. Imagine now that a much smaller amount of fluid is being injected, say $Q\sim1$ [l/s]. This is unlikely the case of a hydraulic stimulation but could instead occur in the case of a natural source of fluids occurring at the same depth. The intensity of the injection (equation \eqref{eq:p*}) is now $\Delta p_*\approx 0.8$ [MPa], and the corresponding stress-injection parameter $\mathcal{T}\approx 2.1$. This will lead equivalently to a rupture that lags the fluid pressure front during continuous injection and that arrest almost immediately after shut-in. The latter, despite the fact that the fault was originally "close" to fail under ambient conditions. On the other hand, a much more intense injection than $\Delta p_*\sim 24$ [MPa] (for instance, with $Q\sim100$ [l/s] and thus $\Delta p_*\sim 80$ [MPa]) will act in the direction of moving the fault response towards the critically-stressed regime (decreasing values of $\mathcal{T}$). Such very intense injection might however likely open the fault, a mechanism that is not accounted in this work. We highlight that according to equation \eqref{eq:p*}, $\Delta p_*$ can be also modified by changes in the fluid dynamic viscosity $\eta$ and fault hydraulic transmissivity $kw$.

Finally, it is worth mentioning that in our model the fluid source is of infinitesimal size. Such idealization provides a proper finite volume of injected fluid $V=Qt_s$, but infinite pressure at the injection point. As a result, fault slip is activated immediately upon the start of injection, irrespective of the injection rate. 
In this regard, our solution must be understood as the late-time asymptotic solution of a more general model where the fluid source possess a finite characteristic lengthscale $\ell_*$. Our results would similarly apply if $t_s$ is much larger than the characteristic diffusion time $\ell_*^2/\alpha$ over the finite source lengthscale. This is indeed the case of most borehole fluid injections, where the borehole radius is $\ell_*\sim 10$ [cm]. By assuming values of hydraulic diffusivity in the range $10^{-4}$ to $1$ [m$^{2}$/s], the characteristic time $\ell_*^2/\alpha$ takes values between $100$ down to $0.01$ [s] which are much smaller than typical fluid injection duration in geo-energy applications.

\subsection{Seismicity driven by post-injection aseismic slip}
We now investigate to which extent post-injection aseismic slip can be considered as a mechanism for the delayed triggering of seismicity. Let us first note that any model that aims at explaining observations of post-injection seismicity must produce spatio-temporal changes of pore-fluid pressure or solid stresses after shut-in. Our model produces both of them. In fact, the pore pressure changes alone are essentially equal to the ones already considered by Parotidis \textit{et al}. \cite{Parotidis_Shapiro_2004} to explain the so-called back front of seismicity that is sometimes observed after the termination of fluid injections (see the inset of figure \ref{fig:on-fault-seismicity}a). Note that Parotidis \textit{et al}.'s model assumes the increase of pore-fluid pressure as the only triggering mechanism of seismicity. We, instead, elaborate on a mechanism based on the combined effect of transient changes in solid stresses due to aseismic slip and pore pressure after shut-in. For the sake of simplicity, we carry out the discussion in terms of circular ruptures only. Rupture non-circularity does not modify the order of magnitude of the results.

\subsubsection{Conceptual model of seismicity}
Let us first define conceptually what seismicity means in the context of our model. We consider that a single fault plane undergoes some transient fluid injection that may be approximated as a pulse of injection rate (see figure \ref{fig:model-schematics}c). It is assumed that as a result of the injection, the fault slides predominantly aseismically. Seismic events are thought to be triggered on unstable patches of the same fault plane (due to, for instance, heterogeneities in rock frictional properties) or other pre-existing discontinuities in  the surroundings of this slowly propagating rupture. The former and latter events are commonly denominated as on-fault and off-fault seismicity, respectively, a terminology that we use later on. Note that an important underlying assumption of this conceptual model is that the unstable patches of the \textit{aseismic} fault do not represent a sufficiently large area to change its predominantly stable mode of sliding to a large dynamic rupture.

We define a Mohr-Coulomb failure function in the form $F\left(\boldsymbol{x},t;\boldsymbol{n}\right)=\left|\tau\left(\boldsymbol{x},t;\boldsymbol{n}\right)\right|-f\left(\sigma\left(\boldsymbol{x},t;\boldsymbol{n}\right)-p\left(\boldsymbol{x},t\right)\right)$, where $\tau$ and $\sigma$ are the shear and total normal stress acting on a certain unstable patch with unit normal vector $\boldsymbol{n}$ at a given position $\boldsymbol{x}$ and time $t$, $p$ is the pore-fluid pressure, and $f$ is a constant (static) friction coefficient. The failure function is such that $F\leq0$ always. The inequality $F<0$ holds when no frictional failure occurs, whereas the equality $F=0$ is valid whenever frictional sliding is activated. The initial stress state is assumed such that $F\left(\boldsymbol{x},t<0;\boldsymbol{n}\right)<0$ at all pre-existing discontinuities before injection starts. Once injection begins, the failure function may approach zero by changes in the solid stresses $\tau$ and $\sigma$ and the pore-fluid pressure $p$. Upon failure of a certain patch, slip will evolve seismically only if during sliding the friction coefficient is allowed to weaken. The time-dependent nucleation of instabilities and, moreover, the explicit modeling of seismicity, are out of the scope of our study. Instead, we consider a conceptual model in which any positive change of the failure function at a certain position and time might lead to frictional sliding and, subsequently, to the \textit{possibility} of triggering an instability.

We introduce two types of changes of the failure function that will prove to be useful for our analysis. The first one, a static change $\Delta F$ between two times, and the second one, an instantaneous change $\dot{F}$:
\begin{equation}
    \Delta F\left(\boldsymbol{x};\boldsymbol{n}\right)=\Delta \tau\left(\boldsymbol{x};\boldsymbol{n}\right)-f\left( \Delta \sigma\left(\boldsymbol{x};\boldsymbol{n}\right)-\Delta p\left(\boldsymbol{x};\boldsymbol{n}\right) \right),\;\text{and}
    \label{eq:yield-function-change}
\end{equation}
\begin{equation}
    \dot{F}\left(\boldsymbol{x},t;\boldsymbol{n}\right)=\dot{\tau}\left(\boldsymbol{x},t;\boldsymbol{n}\right)\textrm{sgn}\left(\tau\left(\boldsymbol{x},t;\boldsymbol{n}\right)\right)-f\left(\dot{\sigma}\left(\boldsymbol{x},t;\boldsymbol{n}\right)-\dot{p}\left(\boldsymbol{x},t\right)\right).
    \label{eq:yield-function-rate}
\end{equation}
$\Delta F$ is widely known in seismology as the Coulomb stress change \cite{King_Stein_1994}, a terminology we adopt hereafter. Note that in equation \eqref{eq:yield-function-change}, we assume that the change of shear stress $\Delta\tau$ is positive if it occurs in the same direction than the shear stress at the selected initial time. On the other hand, in equation \eqref{eq:yield-function-rate}, $\textrm{sgn}\left(\cdot\right)$ is the sign function and the dot over the scalar fields represents a partial derivative in time. $\dot{F}$ is commonly denominated as the Coulomb stressing rate, a quantity that is sometimes correlated to seismicity rates \cite{Dieterich_1994}.

Finally, we highlight that positive contributions to failure $\Delta F>0$ ($\dot{F}>0$) are given by both positive changes of pore-fluid pressure $\Delta p>0$ ($\dot{p}>0$) and negative changes of total normal stress $\Delta \sigma<0$ ($\dot{\sigma}<0$). The effect of the shear stress is however less straightforward and knowledge about the direction of $\tau$ and thus about the absolute state of stress is required to evaluate its relative contribution to $\Delta F$ ($\dot{F}$).

\subsubsection{On-fault seismicity}
On fault, both aseismic-slip stress transfer and pore pressure changes are active during the post-injection stage. We use the Coulomb stressing rate $\dot{F}$ as an indicator of the regions where seismicity is expected. Let us first note that due to the planarity of the fault, the total normal stress rate $\dot{\sigma}$ is zero. Hence, the only component of stress rate that is active along the fault plane is the shear one $\dot{\tau}$. Moreover, since the absolute state of stress of the fault is known in our model, we can assume by convention that $\tau>0$ and consequently that positive rates of shear stress are the ones that contribute to frictional failure. Along the fault plane, equation \eqref{eq:yield-function-rate} thus further simplifies to
\begin{equation}
    \dot{F}\left(\boldsymbol{x},t\right)=\dot{\tau}\left(\boldsymbol{x},t\right)+f\Delta \dot{p}\left(\boldsymbol{x},t\right).
    \label{eq:F-dot-on-fault}
\end{equation}
We first analyze the stress-transfer effect, $\dot{\tau}$ in equation \eqref{eq:F-dot-on-fault}. It is indeed readily to show that the shear stress rate acting on possible unstable patches is positive everywhere along the fault plane. 
For the region outside of the pulse, this can be derived from equation \eqref{eq:momentum-rate} and is graphically demonstrated in figure \ref{fig:slip-rate-characteristics}b. From this figure, we notably observe that seismicity should be triggered predominantly in the proximity of both the locking and rupture fronts, as a consequence of the amplification of shear stress rate concentrated near the tips of the aseismic pulses. Regarding the inside of the pulse, our calculated value of $\dot{\tau}$ in figure \ref{fig:slip-rate-characteristics}b, which is simply equal to the fault shear strength rate, is somewhat irrelevant. The actual stress-transfer effect comes rather from considering that unstable patches may be effectively locked and surrounded by aseismic slip. The latter would increasingly load the locked patches in shear until eventually a dynamic event could be triggered. If the failed unstable patches admit multiple loading-unloading cycles within the time scale in which the rupture pulse passes, various instabilities could be nucleated at the same location in the form of repeating earthquakes or multiplets. Such type of repeating earthquakes have been already detected during injection-induced aseismic slip episodes \cite{Bourouis_Bernard_2007}.

With regard to the pore-pressure effect, $f\Delta\dot{p}$ in equation \eqref{eq:F-dot-on-fault}, its contribution to seismicity can be understood readily from the mere definition of the pore-pressure back front. At distances $r>P(t)$, where $P(t)$ is the radius of the front, the pore pressure rate is positive. Therefore, the region ahead of the pore-pressure back front is where fluid-induced instabilities are expected, a result that was already introduced by Parotidis \textit{et al}. \cite{Parotidis_Shapiro_2004}. Conversely, at distances $r<P(t)$, the opposite holds $\Delta \dot{p}<0$, and thus the triggering of instabilities is here inhibited. 

\begin{figure}
    \centering
    \includegraphics[width=12.9cm]{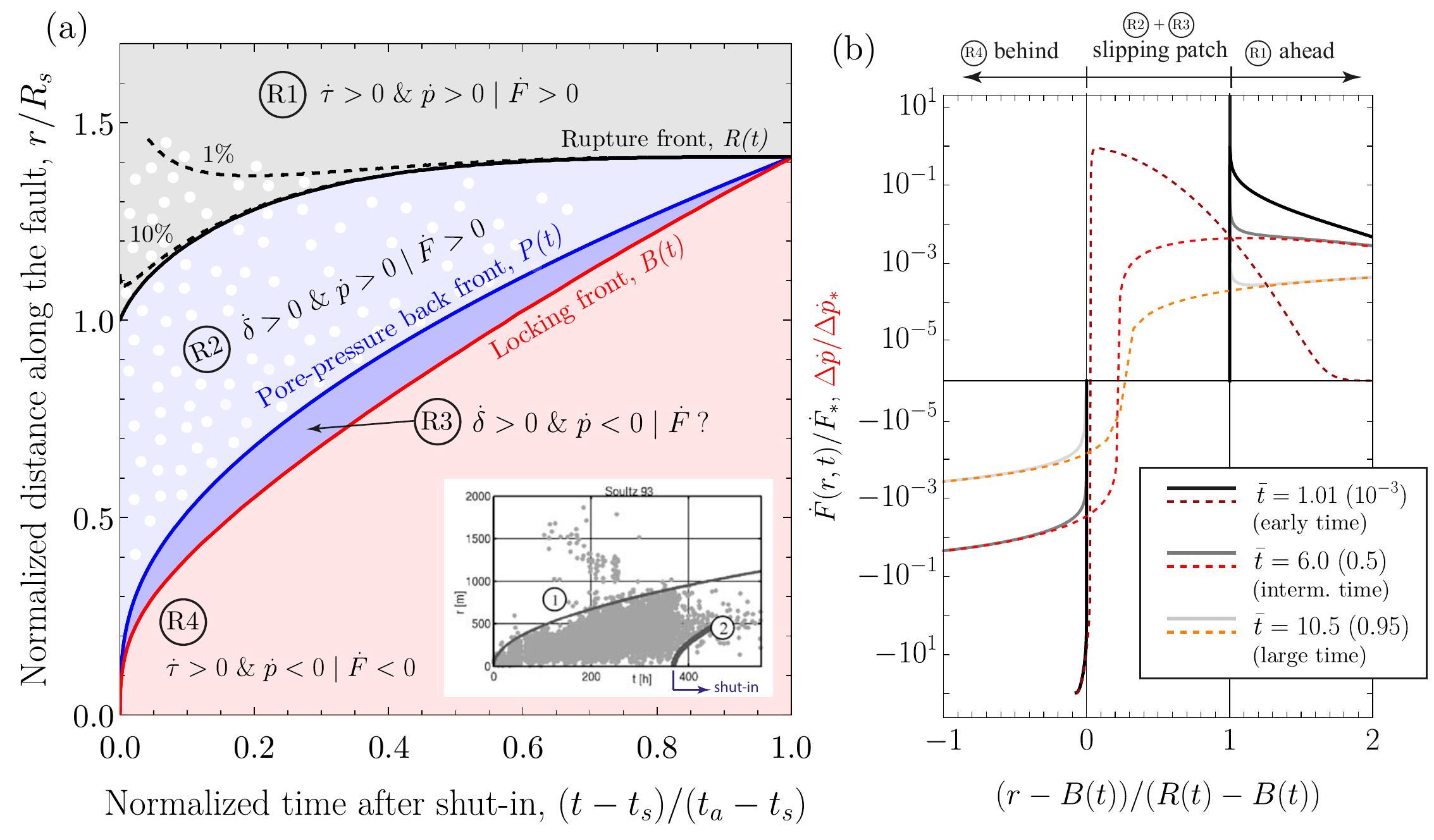}
    \caption{On-fault seismicity ($\mathcal{T}=0.1$ case). (a) Distance versus time plot showing regions over the fault plane where the combined effect of aseismic-slip stress transfer ($\dot{\tau}$ or $\dot{\delta}$) and pore pressure changes ($\dot{p}$) operates differently. White points represent the location and time of possible seismic events in regions where the Coulomb stressing rate $\dot{F}$ is positive. Dashed curves correspond to the position where $\dot{F}$ is 1\% and 10\% of the characteristic Coulomb stressing rate $\dot{F}_*=f\Delta p_*/t_s$. In the inset, distance versus time plot showing the location/time of events during a borehole fluid injection field case, where the so-called back front of seismicity denoted as "2" is observed (modified from \cite{Parotidis_Shapiro_2004}). (b) Normalized Coulomb stressing rate $\dot{F}$ (solid lines in gray scale) and pore pressure rate $\Delta \dot{p}$ (dashed lines in red scale) along the fault plane at three dimensionless times for the same rupture than in (a). Dimensionless times in the legend box are expressed in the format $\bar{t}=t/t_s,\;((t-t_a)/(t_a-t_s))$.}
    \label{fig:on-fault-seismicity}
\end{figure}

We are now ready to superimpose the effects of aseismic-slip stress transfer and pore-pressure diffusion and, notably, to determine the sign of $\dot{F}$ and the spatial distribution of its magnitude. This is partially schematized in figure \ref{fig:on-fault-seismicity}a, where four different regions in which the results of the superposition operate differently, are delineated by the three relevant fronts of the problem, namely, the rupture front $R(t)$, the pore-pressure back front $P(t)$, and the locking front $B(t)$. 
In region R1 for distances ahead of the pulse $r>R(t)$, the Coulomb stressing rate $\dot{F}$ (equation \eqref{eq:F-dot-on-fault}) is always positive. Hence, in this region, seismicity is expected to be triggered by the contribution of both a positive shear stress rate $\dot{\tau}>0$ and a positive pore pressure rate $\Delta\dot{p}>0$. Let us further analyze the magnitude and distribution of $\dot{F}$ here. This is shown in figure \ref{fig:on-fault-seismicity}b, where the spatial profile of normalized Coulomb stressing rate is plotted at various dimensionless times for an exemplifying case with $\mathcal{T}=0.1$. The solitary contribution of the pore pressure rate, which is independent of the value of $\mathcal{T}$, is also included in this figure. Note that the scale of the Coulomb stressing rate $\dot{F}_*$ comes from equation \eqref{eq:F-dot-on-fault} and is equal to $\dot{F}_*=f\Delta p_*/t_s$. We observe from figure \ref{fig:on-fault-seismicity}b that ahead of the rupture pulse and at early times after shut-in, $\dot{F}$ is dominated by the amplification of shear stress rate ($\dot{\tau}\gg\Delta\dot{p}$), whereas at intermediate times and notably at large times, the pore pressure rate becomes the most dominant quantity ($\Delta\dot{p}\gg\dot{\tau}$) from some distance ahead of the rupture front that gets increasingly closer to the slip front with time. However, at the rupture front itself and very close to it, the shear stress rate will always dominate the magnitude of the Coulomb stressing rate due to the square-root singularity of $\dot{\tau}$ discussed in section \ref{slip-rate-stress-rate}. 

Let us now consider the regions R2 and R3 that compose the slipping patch. Here, seismic events are promoted by the increase of shear stress $\dot{\tau}>0$ acting on unstable locked patches that are loaded by surrounding aseismic slip, a mechanism that is denoted by $\dot{\delta}>0$ in figure \ref{fig:on-fault-seismicity}a. Moreover, in region R2 ahead of the pore pressure back front $r>P(t)$, the pore pressure rate is also positive, meaning that the Coulomb stressing rate $\dot{F}$ is strictly positive too. The triggering of instabilities in region R2 is thus expected. Conversely, in the region R3 behind the pore pressure back front $r<P(t)$, the pore pressure rate is negative. As such, its effect counterbalances the positive contribution due to aseismic slip ($\dot{\delta}>0$). Weather $\dot{F}$ is positive or negative in this region is not possible to know from our calculations, because we do not explicitly model the effect of any potential unstable locked patches on the shear loading. Nevertheless, by assuming that the shear loading $\dot{\tau}$ is some unknown but positive quantity, the solitary effect of the pore pressure changes provides a lower bound for $\dot{F}$ within the slipping patch. Even more, the pore-pressure effect $f\Delta\dot{p}$ is a lower bound of $\dot{F}$ over the entire fault plane since as already discussed, the stress-transfer effect $\dot{\tau}$ is positive everywhere. This can be clearly observed in figure \ref{fig:on-fault-seismicity}b when looking at the regions behind and ahead of the rupture pulse.

Finally, in the region R4 behind the slip pulse $r<B(t)$, seismicity is promoted by the amplification of the shear stress rate near the locking front, albeit such amplification is going to be neutralized to some extent by the negative pore pressure rate operating in this region. In fact, figure \ref{fig:on-fault-seismicity}b shows that the negative pore pressure rate appears to completely neutralize the stress rate amplification behind the locking front for the particular case shown in this figure. Furthermore, a general proof of this numerical observation can be developed as follows. First, we recall that $\dot{F}=0$ within the slipping patch and, particularly, when $r\to B(t)^+$. Note that we have just referred to $\dot{F}$ with the same notation than the Coulomb stressing rate acting on unstable locked patches (equation \eqref{eq:F-dot-on-fault}). However, we are rather referring by $\dot{F}$ in this particular case to the rate of the failure function on the portion of the slipping patch that is aseismically sliding. Since $\dot{\tau}$ and $\Delta\dot{p}$ are both continuous at the locking front (the former due to the healing condition discussed in section \ref{evolution-fronts}), the previous limit is also valid when approaching $B(t)$ from the outside of the pulse: $\dot{F}=0$ when $r\to B(t)^-$. Because both $\dot{\tau}$ and $\Delta\dot{p}$ decrease monotonically with increasing distances measured from the locking front and towards the injection point $r=0$, we conclude that $\dot{F}<0$ for all points located behind the locking front, $r<B(t)$. We highlight that this statement is valid at any time after shut-in and for any value of $\mathcal{T}$. More importantly, it allows us to establish unequivocally that seismicity is not expected to occur behind the locking front, despite the positive instantaneous increase of shear stress operating behind the rupture pulse.

The final region along the slip plane where seismicity is expected after shut-in is highlighted in figure \ref{fig:on-fault-seismicity}a by the white points that represent the location and time of possible seismic events, in a similar fashion to the actual data from Parotidis \textit{et al}. \cite{Parotidis_Shapiro_2004} that is displayed in the inset. The lower limit of the active region or, as denominated by Parotidis \textit{et al}. \cite{Parotidis_Shapiro_2004}, the back front of seismicity, is given essentially by the same pore-pressure back front $P(t)$ (equation \eqref{eq:P-front}) introduced first by the previous authors. Weather some instabilities can be triggered in region R3 or not seems irrelevant from an observational point of view, since such small spatial differences are very likely indistinguishable in seismic catalogs and in many cases, possibly even smaller than their location errors. On the other hand, unlike the back front of seismicity that corresponds to a sharp front ($\Delta\dot{p}=0$), the upper limit of the seismically active region or seismicity front is not sharply defined. This is because the Coulomb stressing rate $\dot{F}$ vanishes theoretically only at infinity. The definition of a seismicity front thus requires some degree of arbitrariness likely associated with a chosen threshold to trigger instabilities. In our case, to be somehow consistent with the definition of the back front in terms of rates, one possible choice is to define the front of seismicity as a small percentage of the Coulomb stressing rate undergone in the proximity of the rupture front. Since the scale $\dot{F}_*=f\Delta p_*/t_s$ represents such a quantity at least at early times, we calculate and display curves associated with 1\% and 10\% of $\dot{F}_*$ in figure \ref{fig:on-fault-seismicity}a. We recognize that other choices are possible, notably in terms of Coulomb stress rather than in terms of rates.
 
Finally, note that we also schematize a decrease of seismicity rate with time by drawing increasingly less events (white points) towards right in figure \ref{fig:on-fault-seismicity}a. This is due to the Coulomb stressing rate decreases many orders of magnitude at intermediate and large times comparing to early times (see figure \ref{fig:on-fault-seismicity}b). To provide an understanding of some possible dimensional values of Coulomb stressing rate, let us consider the characteristic values of the same hydraulic stimulation associated with a deep geothermal reservoir introduced in section \ref{critically-stressed-versus-marginally-pressurized}. This example owns an injection intensity $\Delta p_*= 24$ [MPa], injection duration $t_s=1$ [day], and friction coefficient $f=0.6$. Hence, the scale of the Coulomb stressing rate is $\dot{F}_*\approx165$ [Pa/s]. As already mentioned, $\dot{F}_*$ is in the order of magnitude of the Coulomb stressing rate near the rupture front at early times.  $\dot{F}_*$ also corresponds to the scale of the pore-pressure effect alone and thus to the lower bound of $\dot{F}$ within the slipping patch. At intermediate times, the Coulomb stressing rate near the rupture front is $\sim 10^{-2}\times \dot{F}_*$ $\approx1.7$ [Pa/s] (taking the results of figure \ref{fig:on-fault-seismicity}b as reference), whereas at large times $\dot{F}\sim 10^{-4}\times\dot{F}_*$ $\approx0.02$ [Pa/s]. Note that $\dot{F}_*$ does not depend on the stress-injection parameter $\mathcal{T}$, however, the decay of the Coulomb stressing rate with time does depend on $\mathcal{T}$ as the stress-transfer effect does. Considering that the shear stress rate must be proportional to the peak slip rate of the aseismic pulses in an overall sense, figure \ref{fig:slip-rate-characteristics}c can be used as a proxy for understanding the strong dependence of the temporal decay of Coulomb stressing rate with $\mathcal{T}$. Moreover, the early-time and intermediate-time Coulomb stressing rates of this ongoing example are many orders of magnitude above typical values of stressing rate associated with tectonic forces as inferred, for instance, from some major strike-slip faults, $10^{-5}$ to $10^{-3}$ [Pa/s] \cite{Gross_Kisslinger_1997,Parsons_2006}.

\subsubsection{Off-fault seismicity}
\begin{figure}
    \centering
    \includegraphics[width=13cm]{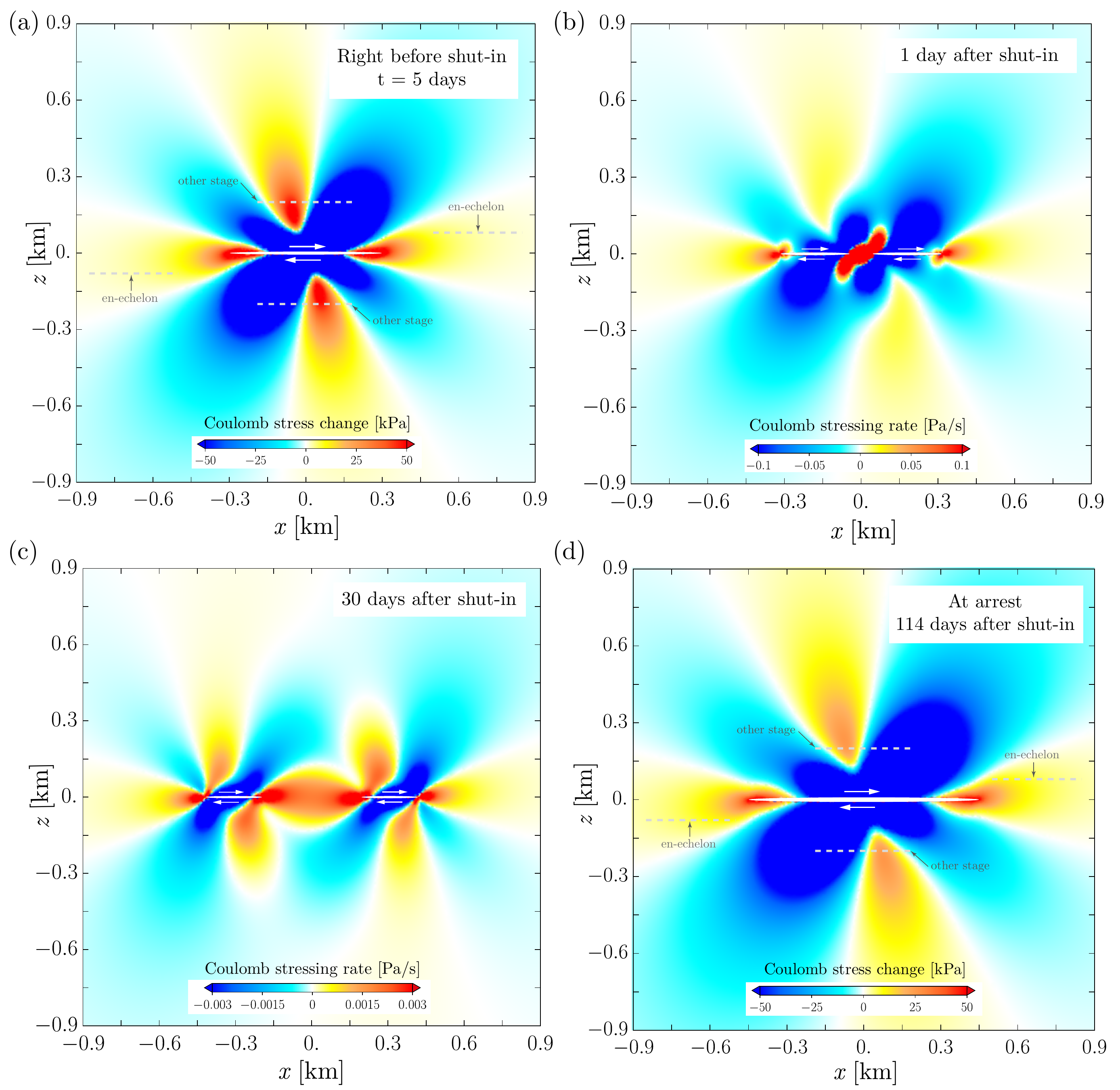}
    \caption{(a) and (d) Coulomb stress change $\Delta F$ and (b) and (c) Coulomb stressing rate $\dot{F}$ on pre-existing discontinuities that are sub-parallel to a reactivated fault lying on the plane $z=0$. This example corresponds to the hydraulic stimulation of a single planar fault in a deep geothermal reservoir, where the stress-injection parameter $\mathcal{T}=0.05$ and $\nu=0$. (a) and (d), $\Delta F$
    from the initial (pre-injection) stress state at $t=0$ to the (a) moment right before stopping the injection, and (d) to the moment in which the rupture pulse arrests. Gray dashed lines schematized sub-parallel fractures associated with either other stimulation stages or en-echelon fracture systems. (b) and (d), $\dot{F}$ after (b) 1 day and (c) 30 days since the start of the injection.}
    \label{fig:off-fault-seismicity}
\end{figure}
Seismicity can be triggered not only on unstable patches laying along the \textit{aseismically} sliding fault, but also on similar patches present in other pre-existing discontinuities nearby the propagating slow rupture. Because fluid flow is assumed to occur only within the permeable fault zone that hosts the reactivated slip plane, off-fault seismicity can be triggered uniquely by the mechanism of stress transfer due to aseismic slip. Equations \eqref{eq:yield-function-change} and \eqref{eq:yield-function-rate} thus take the following forms that exclude the pore pressure changes, $\Delta F\left(\boldsymbol{x};\boldsymbol{n}\right)=\Delta \tau\left(\boldsymbol{x};\boldsymbol{n}\right)-f\Delta \sigma\left(\boldsymbol{x};\boldsymbol{n}\right)$ and $\dot{F}\left(\boldsymbol{x},t;\boldsymbol{n}\right)=\dot{\tau}\left(\boldsymbol{x},t;\boldsymbol{n}\right)\textrm{sgn}\left(\tau\left(\boldsymbol{x},t;\boldsymbol{n}\right)\right)-f\dot{\sigma}\left(\boldsymbol{x},t;\boldsymbol{n}\right)$, respectively. To evaluate the previous expressions, knowledge about the normal vectors $\boldsymbol{n}$ of the target pre-existing discontinuities is required. This information is obviously site specific and always partly uncertain. To keep the analysis general to some extent, there are at least two natural choices for the orientation of discontinuities that seem worth to consider. The first one is the case of discontinuities optimally oriented with regard to stress field: at angles $\theta_*=\pm \left(\pi/4-\arctan(f)/2 \right)$ measured from the local direction of the maximum principal stress. The calculation of the latter direction requires the full initial stress tensor plus computations of its variation in time due to the redistribution of stresses caused by the propagation of the aseismic pulses. Positive changes of the Coulomb stress function would highlight the regions where seismicity is most likely expected in a context in which the orientations of the discontinuities nearby the rupture are fairly unknown. The second option is to consider discontinuities that are parallel or sub-parallel to the reactivated fault. For example, in the context of hydraulic stimulation of deep geothermal reservoirs, multi-stage stimulation techniques 
aim at reactivating isolated sub-parallel fractures that intersect the open-hole section of the stimulation well at different depths/stages. Positive changes of the Coulomb stress function in this case would highlight the effect of one stage on the others. In addition, it is not rare that the reactivated fractures are part of en-echelon fracture systems such that other sub-parallel discontinuities either overlap or form a step-like feature with the reactivated one. Because of its simplicity and relevance for geo-energy applications, we only consider hereafter sub-parallel discontinuities.

Figure \ref{fig:off-fault-seismicity} displays some static and instantaneous changes of the Coulomb stress function for a circular rupture that propagates on a critically-stressed fault with $\mathcal{T}=0.05$. The results are plotted on the $x$-$z$ plane of our global reference system (see figure \ref{fig:model-schematics}) corresponding to the plane where the direction of slip is contained entirely. All calculations in this figure are for a particular choice of parameters representing the hydro-shearing treatment of a deep geothermal reservoir at approximately 4 km depth: injection rate $Q=20$ [l/s], injection duration $t_s=5$ [days], hydraulic diffusivity $\alpha=5\times10^{-3}$ [$\text{m}^{2}/\text{s}$], hydraulic transmissivity $kw=10^{-13}$ [$\text{m}^{2}$], fluid dynamic viscosity $\eta=2\times10^{-4}$ [Pa$\cdot$s], rock shear modulus $\mu=30$ [GPa], constant friction coefficient $f=0.6$, initial total normal stress $\sigma_0=120$ [MPa], initial pore pressure $p_0=40$ [MPa], and initial shear stress $\tau_0=47.9045$ [MPa]. For this particular choice of parameters, the accumulated amount of slip at the injection point at the shut-in time is $\delta\approx1.9$ [cm]. Note that $\delta$ can be approximated analytically from the slip scale $\delta_*$ in the critically-stressed regime, equation \eqref{eq:slip-scale}, with a pre-factor 3.5 derived in equation (28) of \cite{Saez_Lecampion_2022}. 

Figure \ref{fig:off-fault-seismicity}a shows the static change of Coulomb stress $\Delta F$ from the initial conditions to the moment right before the end of the injection. At this time, the rupture is still propagating in crack-like mode, with a radius equal to $R_s=\lambda\sqrt{4\alpha t_s}\approx297$ [m] ($\lambda\approx1/\sqrt{2\mathcal{T}}\approx3.2$). The spatial pattern of $\Delta F$ is remarkably different to the classic pattern of sub-parallel faults having a more homogeneously distributed slip (see figure 2a in \cite{King_Stein_1994} for example) compared to our slip distribution which is more peaked near the injection point (see figure \ref{fig:transition}a). Particularly, we obtain positive changes of Coulomb stress near the center of the rupture, which may affect sub-parallel discontinuities that correspond to other stimulation stages. Such sub-parallel fractures are schematized in figure \ref{fig:off-fault-seismicity}a by dashed lines. In addition, the positive changes of Coulomb stress in the proximity of the rupture front may induce seismicity on nearby discontinuities composing, for instance, an en-echelon arrangement of fractures, a situation that is also schematized in figure \ref{fig:off-fault-seismicity}a.

On the other hand, figures \ref{fig:off-fault-seismicity}b and \ref{fig:off-fault-seismicity}c display the Coulomb stressing rate $\dot{F}$ caused by  the aseismic pulse propagation respectively 1 day and 30 days after the end of the injection. We can observe here the off-fault counterpart of the along-fault amplification of shear stress rate near the locking and rupture fronts shown previously in figure \ref{fig:slip-rate-characteristics}b. Note that the positive rate of Coulomb stress near the locking front affects a region where, before shut-in, the Coulomb stress has dropped $\Delta F<0$ (see figure \ref{fig:off-fault-seismicity}a). Furthermore, the final change of Coulomb stress $\Delta F$ from the beginning of the injection until the moment in which the rupture ultimately arrest (figure \ref{fig:off-fault-seismicity}d), shows that overall there is a drop of Coulomb stress in the off-fault region affected by the stress-rate amplification associated with the passage of the locking front. The latter suggests that off-fault seismicity is not expected to be seen behind or around the locking front, similarly to the case of on-fault seismicity that was analyzed in the previous section. This observation is of course based on the particular case of pre-existing discontinuities that are parallel to the reactivated fault. In this configuration, the main role of post-injection aseismic slip seems to be the one of continuing inducing seismicity in nearly the same regions of positive Coulomb stress change at the time of shut-in (figure \ref{fig:off-fault-seismicity}a) but at increasingly further distances. Note that for this example, the rupture radius at arrest is $R_a\approx449$ [m], which is $R_a/R_s\approx1.51$ times greater than the rupture radius at shut-in, whereas the time of arrest is $t_a\approx119$ [days], that is $t_a/t_s\approx23.8$ times longer than the injection duration. Finally and more generally speaking, depending on the orientation of the surrounding pre-existing discontinuities and the pre-injection stress field, seismicity can continue to be triggered in regions of positive Coulomb stress change after shut-in with significant time delays and at increasingly further distances, perturbing the stress state of large rock volumes with characteristic sizes that are in the order of the spatial extent of the arrested rupture. For instance, for the example considered here, $R_a/R_s\approx1.51$ and thus the rock volume altered by aseismic slip is in the order of $(1.51)^3=3.4$ times larger at arrest than at shut-in.

\subsection{Possible field evidence for post-injection aseismic slip}
Aseismic slip after the shut-in of stimulation wells seems to 
exist in a few cases of geo-energy projects, notably in relation to observations of post-injection seismicity. Our systematic study of this phenomenon allows us now to put those particular cases in perspective. For instance, a hydraulic stimulation for the development of a deep geothermal reservoir was performed in 2013 in the  crystalline basement of the Rittershoffen geothermal field in Northeastern France \cite{Lengline_Boubacar_2017}. The stimulation lasted for 2 days and was accompanied by swarm-like seismicity that illuminated a single planar structure that was reactivated in shear due to the fluid injection. Upon shut-in, seismicity stopped immediately. However, after 4 days in which no micro-seismic activity was detected, a short-lived second swarm began and developed along a nearby sub-parallel structure likely part of an en-echelon fault system \cite{Lengline_Boubacar_2017}. 
Several hypotheses were discussed by these authors to explain what possibly led the second fault to failure and the 4-days delay of this second seismic swarm. The favoured hypothesis for the failure of the second fault was in fact the stress transfer due to injection-induced aseismic slip associated with the first fault, a situation that resembles the off-fault mechanism for en-echelon systems just discussed in the previous section (see figure \ref{fig:off-fault-seismicity}d). However, the authors discarded post-injection aseismic slip as a mechanism for explaining the delayed triggering of the second swarm. They argued that slip on the first fault would be difficult to promote after shut-in due to fault pressure drops quickly to zero, thus moving the fault interface away from failure \cite{Lengline_Boubacar_2017}. 
In light of our results, this statement is valid only locally at the injection point, where the fault interface re-locks immediately upon the stop of the injection. However, as we have seen, fault pressure keeps increasing away from the injection point after shut-in, which drives the propagation of an interfacial pulse-like frictional rupture that further accumulates slip on the fault. Hence, injection-induced aseismic slip can potentially explain as a unique mechanism both the failure of the second fault and the delayed triggering of the swarm of this particular case study
for which the reactivated fault is actually thought to be critically stressed \cite{Lengline_Boubacar_2017}.  

A second interesting case is a long-lived seismic swarm that persisted for more than 10 months after completion of hydraulic fracturing operations of an hydrocarbon reservoir in western Canada in 2016 \cite{Eyre_Zecevic_2020}. In this case study, the authors did attribute the delayed swarm activity to post-injection aseismic slip. Unlike the off-fault setting of the previous case in France, the configuration here is a clear case of on-fault seismicity. Specifically, aseismic slip is thought to be induced by hydraulic fractures that intersect (and pressurize) frictionally-stable segments of a nearby fault at the reservoir level (shales), which in turn transmits solid stresses to distal segments of the same fault but in carbonate units that are thought to slide unstably \cite{Eyre_Eaton_2019}. One key aspect of the proposed conceptual model for post-injection seismicity in this case, is the assumption of heterogeneities in fault permeability to explain a remarkable characteristic of the swarm, namely, a nearly constant rate of post-injection seismicity \cite{Eyre_Zecevic_2020}. The proposed model assumes that elevated pore pressure remains trapped within the fault after shut-in at the reservoir level, due to the extremely low permeability of the shale formations. These trapped fluids are therefore thought to be responsible for a nearly steady propagation of aseismic slip that could explain the constant rate of seismicity \cite{Eyre_Zecevic_2020}. Although their proposed model slightly differs to our model (in which the fault permeability is  homogeneous), this case study does provide strong evidence for the existence of post-injection aseismic slip and its role in the triggering of seismicity after shut-in. Another case in which aseismic slip might have triggered post-injection seismicity is, as suggested by Cornet \cite{Cornet_2016}, the Basel earthquakes in 2006 in Switzerland \cite{Deichmann_Giardini_2009}. 

Further and possibly more conclusive evidence for post-injection aseismic slip may come from carefully-designed laboratory and/or in-situ experiments of fluid injection. Laboratory experiments where a finite rupture grows along a pre-existing interface \cite{Passelegue_Almakari_2020,Gori_Rubino_2021,Cebry_Ke_2022} are particularly promising to explore the post-injection stage. On the other hand, in-situ experiments where the fault deformation is monitored simultaneously at the injection point and at another point away from it \cite{Cappa_Guglielmi_2022}, may also provide the opportunity to investigate this mechanism when combined with hydro-mechanical modelling that includes the depressurization stage \cite{Larochelle_Lapusta_2021}.

\subsection{Model limitations: solution as an upper bound}

Our model contains the minimal physical ingredients to reproduce post-injection aseismic slip in 3D media. As such, the effects of a number of additional effects remain to be investigated. In particular, we have assumed that fluid flow induces mechanical deformation but not vice versa. It has been however suggested for long time that variations of effective normal stress may induce permeability changes in fractures \cite{Witherspoon_Wang_1980} and faults \cite{Rice_1992}. Also, it is known that frictional slip may be accompanied by dilatant (or contracting) fracture/fault-gouge behavior that would inevitably induce changes in fluid flow and thus in the propagation of fault slip (see \cite{Ciardo_Lecampion_2019} for example). Poroelastic effects in the surrounding medium around the fault may be of first order in some cases (see, for instance, \cite{Heimisson_Liu_2022}). Accounting for a permeable host rock may notably speed up the depressurization of pore-fluid within fault zone after shut-in due to the leak-off of fluid. Consequently, the arrest time and maximum run-out distance of the rupture pulses would likely decrease. In this regard, our model in which the leak-off of fluid into the surrounding medium is neglected would represent an upper bound for the arrest time and maximum run-out distance similarly to the case of the arrest of mode I hydraulic fractures \cite{MoLe21}.

In relation to friction, we consider the simplest description that allows us to produce unconditionally stable fault slip, namely, a constant friction coefficient. However, laboratory-derived friction laws \cite{Dieterich_1979,Ruina_1983} are widely used in the geophysics community to reproduce the entire spectrum of slip velocities of natural earthquakes \cite{Scholz_2019}. Unconditionally stable fault slip may be obtained notably in some regimes of so-called rate-strengthening faults when subjected to continuous fluid sources \cite{Dublanchet_2019,Garagash_2021}. For the post-injection problem, a more complex constitutive friction law would add a finite amount of fracture energy as well as frictional healing. In this regard, our model represents a situation in which the fracture energy spent during propagation is zero and there is no possibility for the friction coefficient to heal with time. The dissipation of a finite amount of fracture energy at the rupture front would resist propagation of the aseismic pulses. Similarly  the healing of the friction coefficient would further contribute to the re-locking process of the fault thus accelerating the propagation of the locking front. Both ingredients are therefore expected to shorten the arrest time and maximum run-out distance of the rupture pulses. Our model thus represents also an upper bound with regard to these two mechanisms. Finally, it is worth mentioning that injection-induced aseismic slip also occur during the quasi-static phase preceding the nucleation of dynamic ruptures  \cite{Garagash_Germanovich_2012,Dublanchet_2019}.

\section{Concluding remarks}
\label{conclusion}
We have provided an in-depth investigation of how post-injection aseismic slip propagates and ultimately arrest in 3D media. Our results provide for the first time a conceptual and quantitative framework that may help to understand various observations and applied problems in geomechanics and geophysics associated with slow ruptures driven by the motion of fluids. Among them, a particularly relevant problem for geo-energy applications is the phenomenon of post-injection seismicity. It has been long recognized that aseismic slip may alter the stress state of large rock volumes during borehole fluid injections \cite{Scotti_Cornet_1994}. Based on our findings, we suggest that aseismic slip may continue stressing even larger and more distant regions after shut-in, during timescales that could span even months for fluid injections of only few days, if the reactivated discontinuity is critically stressed as quantified by the value of the stress-injection parameter $\mathcal{T}$. A number of field cases where post-injection aseismic slip may have triggered seismicity in distal on-fault \cite{Eyre_Zecevic_2020} and off-fault \cite{Lengline_Boubacar_2017} regions provide evidence to a varying extent for this mechanism. Further and possibly more conclusive evidence may potentially come from revisiting some case studies via geomechanical modelling and notably from   laboratory and/or in-situ experiments that can either monitor or inferred the propagation of slow ruptures.

Current efforts to manage the seismic risk associated with subsurface fluid injections such as the so-called traffic light systems might be subjected to important limitations in their effectiveness in light of our results. Traffic light systems work under the tacit assumption that operational measures will become shortly effective in preventing the occurrence of events of larger magnitude than some pre-defined threshold \cite{Baisch_Koch_2019}. Our results suggest that instead, the stressing of increasingly larger rock volumes perturbed by aseismic slip may be persistent, if the chosen operational measure is shutting in the well. Future works should therefore focus on developing physics-based strategies to mitigate the seismic risk associated with post-injection aseismic slip.
\vskip6pt

\enlargethispage{20pt}

\ack{The authors acknowledge discussions with Dmitry Garagash about the tip-asymptotic structure of the slip pulses and the healing condition.}
\aucontribute{A.S. Conceptualization, Methodology, Software, Validation, Formal analysis, Investigation, Writing – original draft, Visualization, Funding acquisition. B.L. Conceptualization, Methodology, Software, Writing – review \& editing, Supervision, Funding acquisition.}
\funding{The results were obtained within the EMOD project (Engineering model for hydraulic stimulation). The EMOD project benefits from a grant (research contract SI/502081-01) and an exploration subsidy (contract number MF-021-GEO-ERK) of the Swiss federal office of energy for the EGS geothermal project in Haute-Sorne, canton of Jura, which is gratefully acknowledged. A.S. was partially funded by the Federal Commission for Scholarships for Foreign Students via the Swiss Government Excellence Scholarship.}


\vskip2pc

\bibliographystyle{RS} 

\bibliography{References} 

\end{document}